\documentclass[conference]{IEEEtran}
\IEEEoverridecommandlockouts

\usepackage{cite}
\usepackage{amsmath,amssymb,amsfonts}
\usepackage{graphicx}
\usepackage{textcomp}
\usepackage{xcolor}
\def\BibTeX{{\rm B\kern-.05em{\sc i\kern-.025em b}\kern-.08em
    T\kern-.1667em\lower.7ex\hbox{E}\kern-.125emX}}

\usepackage[color,secparam=\kappa]{crypto/crypto}
\usepackage{amsmath,amsthm}
\usepackage{thmtools}
\usepackage{thm-restate}
\usepackage{graphicx,multirow,array,enumitem,url,stmaryrd}
\usepackage{braket,multicol,xspace}
\usepackage{algorithmicx}
\usepackage[noend]{algpseudocode}
\usepackage[normalem]{ulem}
\usepackage{caption,subcaption}
\usepackage{tikz}
\usepackage{stmaryrd}
\usepackage{bbm}
\usepackage{stmaryrd}
\usepackage[compact]{titlesec}
\usepackage{xcolor}
\usepackage{multirow}
\usepackage{algorithm}
\usepackage{tikz}
\usetikzlibrary{arrows.meta}

\algrenewcommand\algorithmicrequire{\textbf{Input:}}
\algrenewcommand\algorithmicensure{\textbf{Output:}}

\newcommand{\method}[0]{\textsc{SecDVA}\xspace}

\newtheorem{definition}{Definition}[section]
\newtheorem{theorem}{Theorem}[section]

\begin{document}

\title{SecDVA: A Maliciously Secure Multi-Party Framework for Data Value Assessment in Modern Machine Learning Data Markets}


\author{\IEEEauthorblockN{Lixu Wang\IEEEauthorrefmark{1}$^1$,
Xinpeng Yang\IEEEauthorrefmark{1}$^1$\thanks{\IEEEauthorrefmark{1} Equal contribution, \IEEEauthorrefmark{2} corresponding authors.},
Meng Hao$^2$, Guangyi Zhang$^3$, Tianwei Zhang$^1$, Wei Dong\IEEEauthorrefmark{2}$^1$ and
Chenkai Weng\IEEEauthorrefmark{2}$^4$}
\IEEEauthorblockA{$^1$Nanyang Technological University, Singapore, $^2$Singapore Management University, Singapore, \\$^3$Shenzhen Technology University, China, $^4$Arizona State University, The United States}
}

\maketitle

\begin{abstract}
Data Value Assessment (DVA) plays a critical role in modern data markets by enabling buyers to evaluate how a seller’s data would improve model performance before purchase. However, existing DVA approaches require one party to reveal the data or model, exposing participants to privacy leakage, strategic manipulation, and high deployment costs. 
These limitations motivate a DVA protocol that is secure, reliable, widely applicable, and efficient enough for practical deployment in modern data markets.
In this paper, we present \method{}, the first maliciously secure multi-party computation (MPC) framework for reliable and practical DVA in modern machine learning data markets.
\method{} ensures that neither the seller’s data nor the buyer’s model and test data are disclosed beyond the final DVA outcome, while remaining robust against malicious behavior. To address the efficiency and applicability challenges of DVA for large models, we use the neural tangent kernel theory to linearize model updates and introduce lightweight mini-batch approximations to significantly reduce computation overhead. We further design a (2+1)-party MPC protocol based on function secret sharing, together with optimized non-linear primitives. In addition, we optimize the underlying MPC backends, the distributed point function, and the distributed comparison function, reducing the number of Advanced Encryption Standard calls to improve efficiency. Extensive experiments demonstrate that \method{} outperforms existing baselines in the plaintext setting and enables DVA for large-scale models in a reasonable time under MPC. 
\end{abstract}



\section{Introduction}

Data has long been recognized as a critical resource in the information age, and its importance has become increasingly evident with recent breakthroughs in artificial intelligence (AI), as it directly determines AI model performance~\cite{whang2023data}. Besides, with the emergence of large language models (LLMs), the ever-increasing model scale has driven training and data usage costs to unprecedented levels~\cite{amer2023large}. Without carefully assessing whether the data meet the required standards beforehand, blindly investing in model training may not only fail to improve task performance but also result in considerable waste~\cite{miao2024demystifying}. Consequently, data value assessment (DVA) before training is of paramount importance, which has received increasing attention in academia from both the database community~\cite{peng2025reliable, data_market_1} and the AI community~\cite{DVA_review}. Besides, another piece of evidence from industry is that the number of data start-ups, services, and marketplaces~\cite{oceanprotocol2025, harbrdata2025, monda2025} that construct and circulate valuable datasets is steadily increasing. In such contexts, DVA directly influences the pricing, thereby shaping data trading behaviors. 

\begin{figure}
\centering
\includegraphics[width=1.\linewidth]{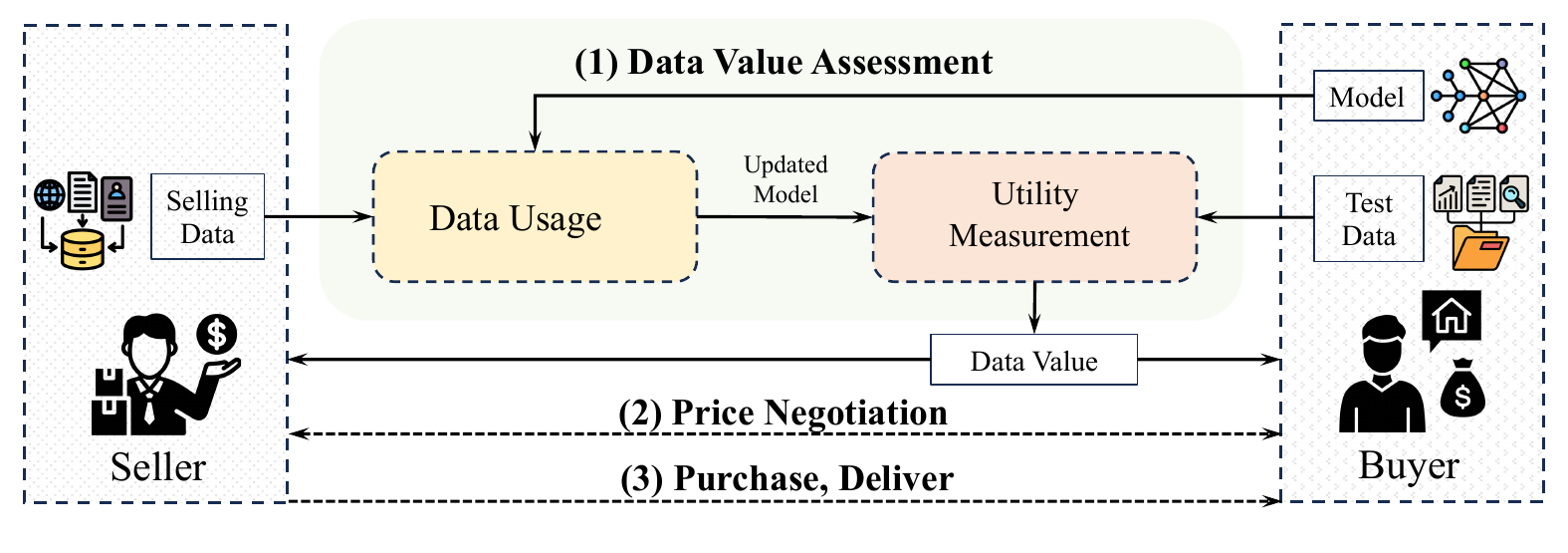}
\vspace{-20pt}
    \caption{End-to-end data trading in modern data markets.}
    \vspace{-10pt}
    \label{fig:fig_1}
\end{figure}

In data trading, many real-world scenarios can be abstracted as a two-party problem~\cite{data_market_2, chen2022selling, agarwal2019marketplace, peng2025reliable}: a buyer seeks to assess how much a seller’s dataset would improve performance on the buyer’s target task, typically defined by the buyer’s model and test data, as shown in Figure~\ref{fig:fig_1}. In practice, such assessments are commonly implemented in one of two ways. First, the buyer sends the target model and test data to the seller, who then returns an assessment result to the buyer~\cite{farrell1996cheap}. Second, the seller provides the dataset to the buyer, enabling the buyer to assess the dataset’s value directly~\cite{chen2022selling}. Both approaches, however, require one party to expose private information to the other. This not only risks commercial and intellectual property loss but also creates room for strategic manipulation, e.g., a seller may inflate DVA results to encourage purchases. Some prior work~\cite{zhou2025ps, shuyue2025efficient} attempts to complete DVA by letting parties perform local computations and exchange only intermediate results. Yet even these intermediate exchanges can leak sensitive information: a buyer may exploit them to (partially) reconstruct the seller’s data~\cite{liu2024vertical}, while a seller may alter its data or computation according to them to inflate the reported value~\cite{naghizadeh2019adversarial, peng2025reliable}.


Overall, four requirements must be satisfied to conduct DVA in modern machine learning (ML) data markets. (i) \emph{Security}: all information related to the properties of the seller and the buyer, except what is revealed by the DVA results, must remain undisclosed. (ii) \emph{Reliability}: DVA results must be produced through reliable computation protocols on verifiable inputs, without any strategic manipulation. (iii) \emph{Applicability}: DVA should apply to real-world data markets, where models are typically large-scale, and discrepancies often exist between the selling data and the buyer’s target task. (iv) \emph{Efficiency}: DVA should be performed efficiently and completed within a reasonable time using limited computational resources.

To build a DVA framework that satisfies these requirements, one feasible approach is to leverage privacy-preserving environments, such as a trusted execution environment (TEE)~\cite{TEE_survey}. However, TEEs require dedicated, expensive hardware, rely on trusting the hardware provider~\cite{TEE_survey}, and suffer from a growing body of side-channel and physical attacks~\cite{TEE_attack_2}. A more practical alternative is secure multi-party computation (MPC), whose security is cryptographically guaranteed. For example, a preliminary study called FullPto~\cite{peng2025reliable} proposes an MPC framework for utility signaling in data markets. Nevertheless, it does not meet DVA requirements in modern data markets. First, it assumes the model is public, which violates the \emph{Security} requirement and allows manipulation via model falsification, undermining \emph{Reliability}. Second, it only supports DVA for regular small-scale models due to \emph{Efficiency} constraints, and it does not allow distributional discrepancies between the seller’s data and the buyer’s target task, which limits \emph{Applicability}. In fact, these limitations reflect the broader challenges of building MPC-based frameworks for modern data markets.


\noindent \textbf{Challenge 1: Tension between Applicability and Efficiency.} Standard DVA consists of two stages, as shown in Figure~\ref{fig:fig_1}. First, the target model is updated using the selling data. Second, the resulting utility gain is measured on the test data. In practice, the data-usage stage typically requires model training. This is manageable for small-scale models, but it becomes prohibitively expensive for large-scale models. Although some work on data valuation~\cite{downsample_1, downsample_2, jia2019efficient, xia2024p} and data selection~\cite{liu2025wasserstein, tv, achille2019task2vec} aims to reduce reliance on full model training, most methods still require at least some amount of training. Moreover, when the training and testing data come from different distributions, DVA often needs additional transfer procedures before utility can be measured~\cite{long2013adaptation}. Implementing any training-involved computation within MPC places substantial pressure on efficiency. For example, even training a small model (with only two convolution and two linear layers) under state-of-the-art semi-honest MPC, whose overhead is much lower than that of maliciously secure MPC, requires over 3.5 hours~\cite{keller2022secure}.


\noindent \textbf{Challenge 2: Tension between Security, Reliability, and Efficiency.}
Most existing protocols~\cite{pang2024bolt,lu2025bumblebee,xu2025breaking} that could potentially support DVA primarily target the semi-honest setting, where parties are assumed to follow the protocol correctly. Achieving security and reliability against malicious deviations requires detecting inconsistencies in both inputs and computation, which introduces substantial overhead. In particular, existing MPC schemes either incur multiple rounds of additional communication by using information-theoretic message authentication codes (MACs)~\cite{gupta2025shark}, or they suffer quadratic computational overhead when relying on distributed zero-knowledge proofs (ZKPs)~\cite{wu2018dizk}. Beyond these generic overheads, DVA also involves many MPC-unfriendly components, especially for large-scale models. Common Transformer operations such as GeLU, LayerNorm, and Softmax are highly non-linear. This prevents directly applying MAC- or ZKP-based techniques that are tailored primarily to linear computations. In addition, DVA typically demands high-precision numerical computation. Existing approaches often rely on low-degree, polynomial-based piecewise approximations, resulting in insufficient accuracy for practical use. Additionally, a malicious data seller may manipulate its dataset to inflate its valuation for specific tasks. Prior work~\cite{peng2025reliable} addresses this issue through dataset authentication to ensure data consistency, but at the cost of substantial computational overhead.

\subsection{Our Proposal}

In this paper, we propose the first MPC framework, \method{}, that enables secure and reliable DVA in modern ML data markets while achieving strong applicability and efficiency by addressing the challenges above. Specifically, to tackle \textbf{Challenge~1}, we leverage recent neural tangent kernel (NTK) theory~\cite{jacot2018neural, lee2019wide, yang2021tensor, malladi2023kernel} to establish a linear relationship between model parameter updates and model outputs, and we use parameter-space projection to estimate model outputs under distribution shift. However, directly applying standard NTK remains computationally expensive. The main bottleneck is constructing the NTK matrix, which requires pairwise inner products between extremely high-dimensional vectors, often on the order of the model’s parameter dimension, across all training samples. To overcome this, we develop a mini-batch strategy to approximate NTK-simulated parameter changes, and we introduce a landmark mechanism to correct the approximation error. With these approximations, \method{} becomes substantially lighter-weight and more MPC-friendly than existing DVA approaches.

To address \textbf{Challenge~2}, we introduce a semi-honest dealer and leverage function secret sharing (FSS)~\cite{boyle2015function} to construct a secure $(2+1)$-party MPC framework, which is substantially more efficient than 2PC used by FullPto~\cite{peng2025reliable}. Under this framework, we implement high-precision non-linear operations, including GeLU, Softmax, LayerNorm, and matrix inversion, to better support costly and precision-sensitive computations in \method{}. To further enhance efficiency, we make non-black-box use of the underlying MPC protocols and ground more optimization to the basic FSS components: the distributed point function (DPF) and the distributed comparison function (DCF), where we propose efficient solutions for maliciously secure FSS evaluation and reduce the overhead of communication and computation. Building upon the improved DPF and DCF, we construct the core MPC gadgets (DReLU and truncation) as well as several intermediate primitives for \method{}, including lookup table, negative exponential, interval lookup, reciprocal, and square-root reciprocal. These components form a highly efficient MPC backend that serves as the foundation of \method{}. Moreover, we compose a committed vector oblivious linear evaluation scheme with the above MPC, which ensures the consistency of the seller's input dataset across the end-to-end data trading. Such input verification is 3 orders of magnitude more efficient than previous ones.

In summary, our contributions include:
\begin{itemize}[leftmargin=*]
    \item We propose the first MPC framework, \method{}, to assess the data value securely for modern ML data markets where data is purchased to tune Transformer models.
    \item We use the neural tangent kernel to make data value assessment lightweight and propose a mini-batch approximation strategy to make the protocol MPC-friendly.
    \item We introduce a semi-honest dealer and adopt FSS to construct a maliciously secure $(2+1)$-party computation setup, optimizing the basic FSS components. Under this setup, we implement non-linear operations used by \method{} on Transformers with high precision and efficiency.
    \item Extensive experiments demonstrate that \method{} substantially outperforms baseline approaches in terms of assessment utility in the plain-text setup, and assesses the value of a data batch on a large transformer model in approximately two hours in the MPC setup, demonstrating its practical feasibility. 
    To demonstrate the efficiency strength of our $(2+1)$-party MPC protocol, we evaluate it on the secure inference task, which is a key component of \method{}. The experiments show that we require only {43.6\%} of the running time and {86.9\%} of the communication compared to the state-of-the-art $(2+1)$-PC protocol.
\end{itemize}





\section{Related Work}

\noindent \textbf{ML Data Markets.} Prior work supports data trading by pricing information and learning utility. Early work studies \emph{arbitrage-free} query pricing, ensuring buyers cannot combine cheaper answers to infer pricier information~\cite{koutris2012query, koutris2013toward, lin2014arbitrage}. More recent mechanisms price predictive capability by selling trained model instances with arbitrage-free guarantees~\cite{chen2019towards}. These designs largely prioritize economic properties (e.g., consistency and revenue) and seldom treat privacy as central. In contrast, Dealer~\cite{liu2021dealer} integrates differential privacy (DP) into both pricing and compensation. Peng et al.~\cite{peng2025reliable} propose a maliciously secure multi-party trading framework to improve reliability and privacy in price signaling, but it assumes a public model; in practice, models are often buyer-specified and should remain private~\cite{agarwal2019marketplace}. Finally, most existing work targets small-scale models rather than today’s large ones.

\smallskip \noindent \textbf{Data Value Assessment.}
Data value is linked to task-agnostic data quality, which is commonly characterized by accuracy, completeness, consistency, timeliness, validity, and uniqueness~\cite{saha2014data}, and is operationalized via constraint-based rules~\cite{chiang2008discovering} and large-scale validation pipelines~\cite{schelter2018automating} to detect anomalies~\cite{aggarwal2001outlier}. Data cleaning improves quality through automated repairs, including probabilistic inference~\cite{rekatsinas2017holoclean} and knowledge-base-driven methods~\cite{chu2015katara}. Quality influences value, but task utility is typically the dominant factor. Two task-aware approaches are widely used: data valuation (DV) and data selection (DS). DV estimates each sample’s marginal contribution using Leave-One-Out~\cite{LOO} or Shapley Value~\cite{sv, ghorbani2019data}, but these are expensive~\cite{luo2024applications, zhang2026efficient}, motivating downsampling~\cite{downsample_1, downsample_2}, training-free estimators~\cite{wu2022davinz, freeshap}, and retraining approximations~\cite{jia2019efficient, xia2024p}. Because DV informs incentives, reliability and security matter: P-Shapley reduces the impact of noisy samples~\cite{xia2024p} and SecSV~\cite{zheng2023secure} secures validation data but leaves training in plaintext. DV also implicitly assumes validation data match the training distribution, which often fails in real-world scenarios. DS instead evaluates transferability before training. Dataset-level distances~\cite{liu2025wasserstein, yang2022fast} are lightweight but require full raw features and may miss task-aware semantics. Model-based DS can better capture task structure via representation embeddings~\cite{achille2019task2vec, tv}, but constructing a model zoo can also be expensive~\cite{nakandala2020cerebro, xing2024database}. These costs are at odds with MPC, motivating lightweight DS methods.

\smallskip \noindent \textbf{Secure Data Usage in AI.}
Recent work develops specialized MPC protocols to use data through either inference or training in AI applications.  
Secure training protocols~\cite{keller2022secure, han2025securexgb} incur prohibitive overhead and do not scale to modern Transformers, so we focus on secure inference.  
Most Transformer protocols assume the semi-honest model.  
Hybrid schemes such as BOLT~\cite{pang2024bolt}, Bumblebee~\cite{lu2025bumblebee}, and BLB~\cite{xu2025breaking} combine homomorphic encryption (HE) and secret sharing (SS), using efficient HE packing for linear layers and SS-friendly piecewise polynomials for nonlinearities like GeLU, Softmax, and LayerNorm.  
Other approaches employ FHE for non-interactive inference~\cite{zhang2025nexus} or secure three-party frameworks~\cite{gupta2024sigma}. 
General-purpose MPC with malicious security~\cite{keller2020mp} is too expensive for Transformers.  
Shark~\cite{gupta2025shark} offers maliciously secure inference with a semi-honest dealer but uses coarse approximations (e.g., ReLU for exponentials, quadratic functions for GeLU) that degrade precision.  
No existing semi-honest or maliciously secure MPC framework fits DVA in modern data markets.

\section{Preliminaries}

\subsection{Data Value Assessment}
\label{subsec:dva}

\begin{definition}[Data Value Assessment]
\label{def:DVA}
Let $\mathcal{D}$ be a dataset to be valued, $\mathcal{T}$ be a benchmark dataset associated with a target task, and $f_\theta$ (parameterized with $\theta$) denote a target model used to solve the task on $\mathcal{T}$. A \emph{data value assessment (DVA)} mechanism is a mapping $\mathsf{V}:\; V(\mathcal{D},\mathcal{T},f_\theta)\ \mapsto\ \nu \in \mathbb{R},$ which quantifies the value of $\mathcal{D}$ in improving the performance of $f_\theta$ on $\mathcal{T}$ under a prescribed evaluation protocol.
\end{definition}
Definition~\ref{def:DVA} is intentionally general: the value $\nu$ depends not only on the raw samples in $\mathcal{D}$, but also on \emph{how} $\mathcal{D}$ is used, the model class and initialized parameters in $f_\theta$, and the evaluation criterion on $\mathcal{T}$. In principle, an ideal assessment would require identifying the \emph{optimal} usage strategy of $\mathcal{D}$ to improve $f_\theta$ on the target task, which is typically intractable. A widely adopted approximation paradigm can be described as a two-stage procedure~\cite{peng2025reliable, agarwal2019marketplace}:

\smallskip \noindent \textbf{Data usage.}
Fix a \emph{data-usage operator}
$\mathsf{O}:\; O(f_\theta,\mathcal{D}) \mapsto f_{\theta_{\mathcal{D}}}$, which specifies how $\mathcal{D}$ is exploited to obtain a derived model $f_{\theta_{\mathcal{D}}}$ from the original model (or initialization) $f_\theta$. The operator $\mathsf{O}$ encapsulates all algorithmic choices, among which the most commonly used is the gradient descent training.

\smallskip \noindent \textbf{Utility measurement.}
Fix a \emph{utility functional} $\mathsf{U}:\; U(\mathcal{T}, f_\theta) \mapsto u \in \mathbb{R}$, which evaluates the model utility on the benchmark dataset $\mathcal{T}$ (higher is better), such as accuracy, AUROC, mean IoU, BLEU, or negative loss. The value of $\mathcal{D}$ with respect to $\mathcal{T}$ is then measured by the utility gain relative to the baseline:
\[
V(\mathcal{D}, \mathcal{T}, f_\theta)
\;=\;
U(\mathcal{T}, f_{\theta_{\mathcal{D}}}) - U(\mathcal{T}, f_\theta).
\]
The specific measurement depends critically on the relationship between $\mathcal{D}$ and $\mathcal{T}$. When $\mathcal{D}$ and $\mathcal{T}$ are drawn from similar distributions and correspond to the same (or highly aligned) task, the value can be measured directly by the improvement in the target-task utility on $\mathcal{T}$.
When $\mathcal{D}$ and $\mathcal{T}$ differ substantially, especially when they are associated with different tasks, a direct utility comparison may be misleading unless a concrete transfer protocol is specified~\cite{long2013adaptation, xing2024database}. Alternative transfer characterizations include improvements in few-shot performance, e.g., achieving a higher target accuracy with the same labeled examples from $\mathcal{T}$.

\subsection{Multi-party Computation}
\label{sec:prelim:mpc}

Secure multi-party computation (MPC) is a cryptographic approach that allows multiple parties $P_0, P_1,\dots$ to jointly compute a function $f(\cdot)$ with private inputs $x_0,x_1,\dots$ while learning nothing beyond the output $y = f(x_0,x_1,\dots)$. MPC protocols mainly involve an encoding scheme that supports the secret sharing of private values and a series of protocols that compute functions over secret shares~\cite{micali1987play,yao1986generate,yao1982protocols}.

\shortsection{Secret Sharing.} Given a statistical security parameter $\lambda$ and $\ZZ_{q}$ as a ring of size $q$, two-party additive secret sharing of $x \in \ZZ_{2^n}$ defines one share as a randomly sampled $x_0 \randfrom \ZZ_{2^n}$ and another share as $x_1 \!\eqdef\! x - x_0 \bmod 2^n$. The authenticated secret sharing with information-theoretic message authentication codes (IT-MACs) requires each party to hold a MAC key $\Delta_i \in \ZZ_{2^\lambda}$~\cite{nielsen2012new,damgaard2012multiparty}. 
For each share $x$, party $P_i$ holds $M_i \in \ZZ_{2^{n+\lambda}}$ such that $ M_0 + M_1 = (x_0 + x_1)  \cdot (\Delta_0+\Delta_1) \bmod 2^{n+\lambda}$.
The protocols built upon authenticated secret shares ensure that any malicious behavior is detectable, and honest parties can abort the execution before any secret is revealed. We denote an authenticated secret-shared value $x$ as $\rshare{x}$.

\shortsection{Function Secret Sharing (FSS).} 
Given a function $f(\cdot): \ZZ_{2^n} \rightarrow \GG$, FSS splits it into two functions $(f_0(\cdot), f_1(\cdot))$ such that $f(\cdot) \eqdef f_0(\cdot) + f_1(\cdot)$ and each of the shares is held by one party. We make use of two practical FSS named distributed point function (DPF) and distributed comparison function (DCF)~\cite{boyle2021function}. 
DPF is defined as follows~\cite{ boyle2015function}.

\begin{itemize}[leftmargin=*]
  \item $(k_0,k_1) \from {\sf DPF}.{\sf Gen}(1^\kappa,{\alpha,\beta})$: On input $1^\kappa$, point $\alpha$ and value $\beta$, output a pair of keys $(k_0,k_1)$. We omit $1^\kappa$.
  
  \item $t_b \from {\sf DPF}.{\sf Eval}(k_b, x)$ for $b\in \bool$: On input a DPF key $k_b$ and a point $x \in [0,2^n)$, output an additive secret share of $t$ such that $t_0+t_1 = \beta$ if $x = \alpha$ and $0$ otherwise.
\end{itemize}


DCF shares similar key generation and evaluation as DPF, except that it outputs $t_0+t_1 =\beta$ if $x\geq \alpha$, and $0$ otherwise.

\shortsection{Vector Oblivious Linear Evaluation (VOLE).}
VOLE is a vector extension of oblivious linear evaluation
(OLE)~\cite{applebaum2017secure}.
Subsequent works developed efficient methods for generating
VOLE correlations~\cite{boyle2018compressing,boyle2019efficient}.
Specifically, over a finite field $\mathbb{F}$, VOLE generates
correlated values $({x},{M})$ for $P_0$
and $(\Delta,{K})$ for $P_1$, satisfying ${M}={K}+\Delta{x}$, where ${x},{M},{K}\in\mathbb{F}^m$
and $\Delta\in\mathbb{F}$. This relation, called a VOLE
correlation, provides a MAC-like authentication structure in secure computation.

\shortsection{Maliciously Secure MPC in the Preprocessing Model.} 
To prevent information leakage, MPC is conducted over secret-shared data and supports the following operations. 
\begin{itemize}[leftmargin=*]
\item Linear combination: From $c_0,\dots,c_{\ell}$ and $\rshare{x_1},\dots,\rshare{x_{\ell}}$, compute $\rshare{x} = c_0 + \sum_{i\in [\ell]} c_i \cdot \rshare{x_i}$.
\item Multiplication: From $\rshare{x_0},\rshare{x_1}$, compute $\rshare{x} = \rshare{x_0} \cdot \rshare{x_1}$.
\end{itemize}

MPC in the preprocessing model refers to a preprocessing phase that generates input-independent correlations, and an online phase that executes the input-dependent operations~\cite{beaver1996correlated}. It is generally more efficient than one-shot execution. Taking the multiplication as an example. In preprocessing, parties generate random multiplication triples $(\rshare{a},\rshare{b},\rshare{c})$ where $c=a\cdot b$ for arbitrary random $a$ and $b$. 
During the online phase, $\rshare{z} \eqdef \rshare{x} \cdot \rshare{y}$ can be computed by $\rshare{z} \eqdef c -e \rshare{b} - f \rshare{a} +ef$, where $e \eqdef a -x$ and $f \eqdef b-y$ can be safely opened without revealing secret $(x,y)$.

\shortsection{(2+1)-party Computation.}
The setting introduces a dealer (assumed semi-honest) to participate in the preprocessing. In the previous case, the dealer samples the triple and distributes the shares of $(a,b,c)$ to the other two parties.
It comes with the following advantages. (i) It simplifies the time-consuming triple generation process and accelerates the protocol execution. (ii) The dealer only participates in input-independent preprocessing, so it learns no information as long as it does not collude with other parties in the online phase.


\shortsection{Committed-Input MPC.}
Cryptographic commitment to a data $\vecx$ refers to a tuple of information $({\sf com}, {\sf decom})$ and two functions $({\sf com}, {\sf decom}) \leftarrow {\sf Commit}(\vecx)$ and ${\sf re} \leftarrow {\sf Open}({\sf com}, {\sf decom})$ in which ${\sf re}$ interprets to acceptance or rejection. It needs to satisfy \emph{hiding} and \emph{binding} properties. \emph{Hiding} requires that the commitment ${\sf com}$ does not reveal any information of $\vecx$. \emph{Binding} requires that ${\sf com}$ cannot open to any $\vecx' \neq \vecx$. Committed-input MPC requires that a party $P_0$ commits to $\vecx$ by $({\sf com}_x, {\sf decom}_x) \leftarrow {\sf Commit}(\vecx)$ and publishes ${\sf com}_x$ to $P_1$. Later, it needs to secretly share $[\vecx]$ and prove that $[\vecx]$ is consistent with its commitment ${\sf com}_x$.
\section{Problem Formulation}
\label{sec:formulation}

\subsection{Data Value Assessment in ML Data Markets}
\label{sec:utility-signaling}

In data marketplaces, a buyer $P_1$ often wishes to assess the value of a candidate dataset before purchase, while the seller $P_0$ is unwilling to reveal the raw data due to economic and confidentiality concerns. Concretely, suppose $P_0$ holds a dataset $\mathcal{D}=\{(\bm{x}_i,\bm{y}_i)\}_{i=1}^{N_\mathcal{D}}$ offered for sale on a trading platform $P^*$, and $P_1$ holds a benchmark dataset $\mathcal{T}=\{(\bm{x}_i,\bm{y}_i)\}_{i=1}^{N_\mathcal{T}}$ for evaluation. $P_1$ also has a target model $f_\theta$ whose performance on $\mathcal{T}$ is expected to improve. Then, before making the purchase, $P_1$ aims to assess the value of $\mathcal{D}$ in improving the performance of $f_\theta$ on $\mathcal{T}$; following Definition~\ref{def:DVA}, this value is denoted by $V(\mathcal{D},\mathcal{T},f_\theta)$.

In real-world data trading, however, both $\mathcal{D}$ and $\mathcal{T}$ remain private to the seller and buyer, respectively, before purchase. As a result, the relationship between their underlying data-generating distributions is typically unknown. Let $\mathcal{D}$ be drawn i.i.d.\ from a joint distribution over inputs and targets, $(X_{\mathcal{D}},Y_{\mathcal{D}})\sim P_{\mathcal{D}}$, and let $\mathcal{T}$ be drawn from $(X_{\mathcal{T}},Y_{\mathcal{T}})\sim P_{\mathcal{T}}$. The ideal case is $P_{\mathcal{D}} \equiv P_{\mathcal{T}}$, but this is rare in practice. More commonly, trading scenarios exhibit \emph{feature discrepancy} (mismatched input spaces or measurement processes) and/or \emph{label discrepancy} (mismatched task definitions or annotation schemas). These discrepancies necessitate distinct procedures in DVA as introduced in Section~\ref{subsec:dva}.

\subsection{Privacy Model}
\label{sec:privacy-model}

The above DVA involves three sensitive objects:  $\mathcal{D}$, $\mathcal{T}$, and $f_\theta$. The privacy model should explicitly ensure that they remain private to their respective owners before purchase. The motivation is threefold. First, they may contain private or regulated information (e.g., patient images, EHR time series, or clinical notes in the medical examples), and models can also leak sensitive details through attacks such as extraction or inversion. Second, they embody valuable proprietary assets: both sellers and buyers invest substantial effort, resources, and expertise in constructing datasets and training models, and these assets are often directly tied to core commercial interests. Unauthorized exposure can therefore cause significant economic harm. Third, disclosure of any of these objects can enable value inflation or manipulation, which we analyze in detail in the trust model that follows.

\subsection{Trust Model}
\label{sec:trust-model}


\smallskip \noindent \textbf{Adversarial seller and buyer.}
Either party may attempt to manipulate the protocol to improve its payoff:
\begin{itemize}[leftmargin=*]
    \item \emph{Protocol Deviation.} Parties may not faithfully execute the prescribed mechanism $\mathsf{V}$, e.g., skipping steps, using modified hyperparameters or randomness, selectively dropping samples, or crafting responses inconsistent with the protocol transcript. In particular, a deviating seller aims to inflate or otherwise manipulate the reported value, while a deviating buyer aims to maximize information gained about the seller's dataset (e.g., by inducing additional side outputs or adaptively steering the computation), potentially biasing $\nu$.
    \item \emph{Dataset Falsification.} The seller may present a dataset $\tilde{\mathcal{D}}\neq \mathcal{D}$ (e.g., cherry-picked, duplicated, poisoned) to inflate the reported value. This falsification is easy to achieve, especially when information about $\mathcal{T}$ and $f_\theta$ is disclosed, underscoring the need to protect their privacy.
\end{itemize}

\smallskip \noindent \textbf{Semi-honest platform.}
Many markets operate through a data trading platform that orchestrates protocol execution and message passing. We model this platform as \emph{semi-honest}: it follows the specified computation and communication procedures and does not collude with either other party, but may try to infer additional information from transcripts and intermediate values. This assumption is weaker than a fully trusted third party and stronger than a fully malicious adversary. In practice, this platform is typically fulfilled by legal data-trading companies or large cloud service providers, which have strong incentives to avoid collusion and malicious behaviors that could jeopardize their business reputation.

\section{Methodology}

\subsection{Strawman Maliciously Secure Two-party Protocol}
\label{subsec:strawman-2pc-dva}

\begin{figure}
\centering\includegraphics[width=.9\linewidth]{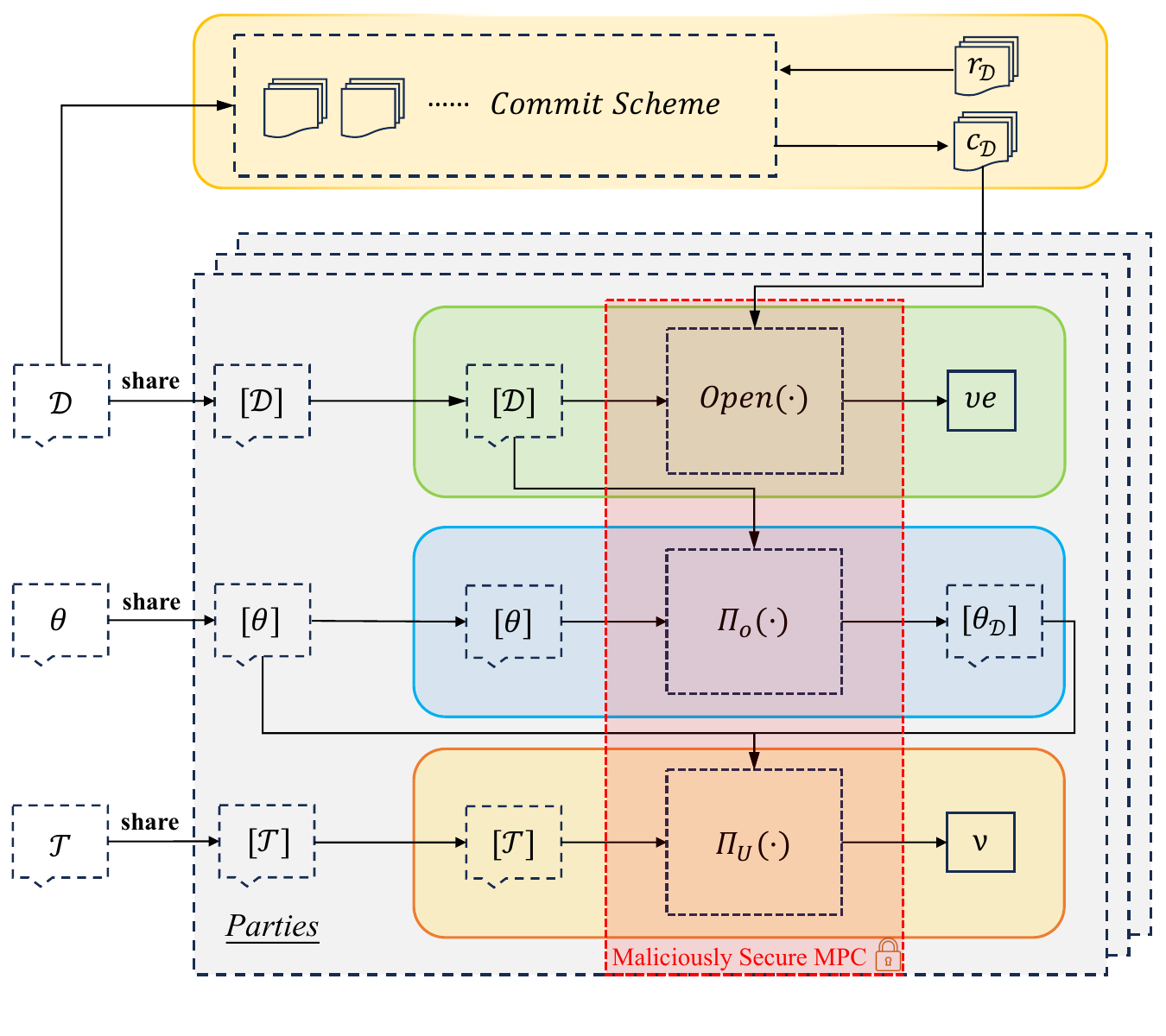}
\vspace{-15pt}
    \caption{The high-level procedure of the strawman maliciously secure two-party protocol for data value assessment.}
    \vspace{-15pt}
    \label{fig:strawman}
\end{figure}

We present a strawman protocol for realizing DVA in a two-party computation (2PC) setting under malicious security, as shown in Figure~\ref{fig:strawman}. The goal is to compute the DVA value $\nu = V(\mathcal{D}, \mathcal{T}, f_\theta)$ while keeping $\mathcal{D}$ private to the seller and $\mathcal{T}, \theta$ private to the buyer, except for what is implied by the output $\nu$. The parties agree on a public specification $\mathsf{V}$ that fixes the data-usage operator $O$, the utility functional $U$, and the concrete implementation details such as preprocessing, numeric format, randomness generation, and the training budget/hyperparameters.
The protocol uses a cryptographic commitment scheme $\mathsf{Com}$ and a malicious-secure 2PC protocol $\Pi$ for evaluating arithmetic and boolean circuits with malicious security. For any operation $O,U$ in the plaintext, we define the corresponding 2-party secure computation protocol as $\Pi_O, \Pi_U$. 
To prevent \emph{Dataset Falsification}, we require the seller to commit their dataset $\mathcal{D}$ in advance while $\Pi$ is employed to detect any \emph{Protocol Deviation} of both the seller and the buyer.

\subsubsection{Protocol Procedure}

\noindent \emph{\textbf{Commit phase.}} The seller commits to its dataset by generating a commitment
$c_{\mathcal{D}} \leftarrow \mathsf{Com}(\mathcal{D}; r_{\mathcal{D}})$ where $r_{\mathcal{D}}$ is a randomness
and sends $c_{\mathcal{D}}$ to the buyer. For example, this commitment
may be instantiated as a cryptographic hash of $\mathcal{D}$ for later
verification. The public DVA specification $\mathsf{V}$ is treated as common input (or, optionally, a committed transcript hash can be exchanged for auditability). Basic well-formedness conditions for $\mathcal{D}$ are enforced by checks inside the secure computation, and deviations lead to abort under the malicious security guarantees of $\Pi$.

\noindent \emph{\textbf{Data usage phase.}} The parties first share their private inputs secretly and obtain $\rshare{f_{\theta}}$, $\rshare{\mathcal{T}}$, and $\rshare{\mathcal{D}}$.
They then jointly compute $\rshare{f_{\theta_{\mathcal{D}}}} \!=\! \Pi_O(\rshare{f_{\theta}}, \rshare{\mathcal{D}})$. The protocol evaluates the agreed operator $O$, which may correspond to a prescribed training procedure (e.g., a fixed number of SGD steps with predetermined hyperparameters and randomness derived from an agreed PRG seed). The resulting model $f_{\theta_{\mathcal{D}}}$ remains secret-shared, ensuring that neither party learns the updated parameters in the clear. To guarantee consistency, the parties verify that the shared dataset $\rshare{\mathcal{D}}$ matches the commitment $c_{\mathcal{D}}$ by computing $ve \leftarrow \mathsf{Open}(\rshare{\mathcal{D}}, c_{\mathcal{D}}, r_{\mathcal{D}})$. The buyer proceeds only if $ve = \mathsf{true}$; otherwise, the protocol aborts. 

\noindent \emph{\textbf{Utility measurement phase.}} Using this 2PC backend, the parties compute the baseline utility $\rshare{u_0} = \Pi_U(\rshare{\mathcal{T}}, \rshare{f_\theta})$ and the post-usage utility $\rshare{u_1} = \Pi_U(\rshare{\mathcal{T}}, \rshare{f_{\theta_{\mathcal{D}}}})$ on the buyer's secret-shared benchmark dataset $\rshare{\mathcal{T}}$. The DVA value is computed as $\rshare{\nu} = \rshare{u_1} - \rshare{u_0}$, and the final $\nu$ is revealed to both parties.

\subsubsection{Challenges.}

\begin{itemize}[leftmargin=*]
    \item Commonly used commitment schemes (e.g., hash-based or public-key-based constructions) incur significant overhead in the above setting, as their verification procedures are often complex and poorly suited for MPC. Although Peng et al.~\cite{peng2025reliable} reduce this overhead by combining Merkle hash trees with a sampling strategy, the resulting approach still introduces non-negligible costs. 
    
    \item The data-usage phase is inherently expensive, as it typically involves model training or fine-tuning. These operations are already computation-intensive in the plaintext setting and become substantially more costly when executed under maliciously secure 2-party computation.
    
    \item Implementing operations within maliciously secure 2PC is itself highly expensive, especially for large-scale models. Core operations such as large matrix multiplications, nonlinear activation functions, normalization layers, softmax computations, and optimizer updates incur significant cryptographic overhead, rendering end-to-end secure training and evaluation impractical in realistic applications.
    
\end{itemize}

\subsection{Our Proposed \method{}}
\noindent \textbf{Overview.} To achieve maliciously secure DVA, we first make the data usage phase as lightweight as possible in plain-text settings. We leverage NTK theory to linearly map the model parameter change and the model output, and use model parameter projection to estimate the model output across different data distributions. However, employing standard NTK also requires expensive operations. Therefore, we develop a mini-batch strategy to approximate standard NTK-simulated model parameter changes and design a landmark mechanism to correct the error of such approximation. 
As for the MPC protocol for \method{}, we design dedicated mechanisms for input verification and private DVA. 
To enable efficient consistency checks over the input data, we leverage committed vector oblivious linear evaluation (C-VOLE)~\cite{sun2025committed}, which is aligned with our (2+1) SPDZ-MPC framework~\cite{damgaard2012multiparty}.
To realize private DVA, we first break down the task into the commonly used model components, such as NTK, MHA, and GeLU. Then, we emulate these functions with more fundamental linear or nonlinear mathematical operations, such as matrix inverse, exponentiation, and reciprocal square root. Finally, to support efficient evaluation of these basic operations with high precision, we build an MPC backend based on (2+1) SPDZ-MPC, combined with our improved function secret sharing.


\subsection{Lightweight DVA Designs}
\label{sec:ntk_training_free}

As introduced in Section~\ref{subsec:dva}, we typically update the model on the seller’s dataset via gradient descent and then evaluate the resulting model utility. Importantly, the utility function depends on the model’s outputs. Thus, the purpose of running gradient descent here is to obtain the change in model outputs induced by training. For large-scale neural networks, simulating gradient descent can be lightweight. Specifically, we can estimate the post-update outputs using an approximation based on the linearized regime, which relates parameter updates to output changes through a first-order expansion. Concretely, we use the theorem,

\begin{theorem}[Gradient Descent Linearization~\cite{lee2019wide}]
Consider a model $f$ with $L$ layers whose parameters construct a set $\theta$. Let $w_1\!=\!\cdots\!=\!w_L\!=\!w$ denote the width of each layer. Suppose the Jacobian on all training inputs is $J(\mathcal{X};\theta) \!=\! \partial f(\mathcal{X};\theta) / \partial \theta$, and the Neural Tangent Kernel (NTK) is $\mathbf \Theta(\mathcal{X}, \mathcal{X};\theta) \!=\! J(\mathcal{X}; \theta) J(\mathcal{X}; \theta)^\top$. Assume the minimum eigenvalue $\lambda_{\min}(\mathbf \Theta)\!>\!0$. Applying gradient descent with learning rate $\eta\!<\!\eta_{\mathrm{critical}}:=2(\lambda_{\min{}} (\mathbf \Theta)+\lambda_{\max{}} (\mathbf \Theta))^{-1}$, for every $\bm{x}\!\in\!\mathbb{R}^{w_0}$ with $\|\bm{x}\|_2\!\le\! 1$, with probability close to $1$ over random initialization when $w\!\to\! \infty$ and at any time $t$, 
\[
\sup_{t\ge 0}\!\bigl\|f_t(x)-\!f_t^{\mathrm{lin}}(x)\bigr\|_2, \sup_{t\ge 0}\!\frac{\|\theta_t\!-\!\theta\|_2}{\sqrt{n}}, \sup_{t\ge 0}\!\bigl\|\hat{\Theta}_t\!-\hat{\Theta}\bigr\|_2 \!=\! \mathcal{O}\!\left(w^{-\frac{1}{2}}\right).
\]
\end{theorem}

Recent works~\cite{yang2021tensor, malladi2023kernel} show that this linearization phenomenon holds broadly across common deep architectures and large pre-trained models. Under this regime, the model output at parameters $\theta_t$ admits the first-order approximation
\begin{equation}
f(\mathcal{X};\theta_t) = f(\mathcal{X};\theta) + J(\mathcal{X};\theta)\cdot(\theta_t-\theta).
\label{eq:linear_model}
\end{equation}

However, this approximation only provides output changes on the training inputs $\mathcal{X}$. Our goal, in contrast, is to estimate utility on the buyer’s dataset $\mathcal{T}$, whose input distribution may differ from the seller’s dataset $\mathcal{D}$. To bridge this mismatch, we project the parameter displacement induced by training on $\mathcal{D}$ onto the displacement direction induced by training on $\mathcal{T}$, and quantify how much of the $\mathcal{T}$-relevant update can be achieved by $\mathcal{D}$ (as there may be label discrepancy, we restrict the projection to the feature extractor parameters). We formalize an ``optimal'' model as follows.

\begin{definition}[Optimal Model]
A model $f$ is optimal (for dataset $\mathcal{D}$ and loss $\mathcal{L}$) if its parameters $\theta$ attain the minimal empirical loss, i.e., $\theta^*\in \arg\min_{\theta}\mathcal{L}(\mathcal{D};\theta)$.
\end{definition}

Let $\theta_\mathcal{D}^*$ and $\theta_\mathcal{T}^*$ denote optimal parameters obtained by training on $\mathcal{D}$ and $\mathcal{T}$, respectively, starting from the same reference point $\theta$. Define the corresponding displacements $\Delta\theta_\mathcal{D}:=\theta_\mathcal{D}^*-\theta, \, \Delta\theta_\mathcal{T}:=\theta_\mathcal{T}^*-\theta$.
We then project $\Delta\theta_\mathcal{D}$ onto the direction of $\Delta\theta_\mathcal{T}$:
\begin{equation}
\Delta \theta_{\mathcal{D}\to\mathcal{T}} := \Delta\theta_\mathcal{T} \langle \Delta\theta_\mathcal{D},\Delta\theta_\mathcal{T}\rangle / \|\Delta\theta_\mathcal{T}\|_2^2,
\label{eq:projection}
\end{equation}
where $\langle\cdot,\cdot\rangle$ is the Euclidean inner product between two vectors. Intuitively, $\Delta \theta_{\mathcal{D}\to\mathcal{T}}$ measures the component of the seller-induced update that aligns with the buyer-relevant update direction. Using the linearized model as shown in Eq.~\eqref{eq:linear_model}, the resulting change in model outputs on $\mathcal{T}$ is proportional to the Jacobian on $\mathcal{T}$ applied to this projected displacement: $V(\mathcal{D},\mathcal{T},\theta) \propto J(\mathcal{X}_\mathcal{T};\theta)\cdot \Delta \theta_{\mathcal{D}\to \mathcal{T}}$.

In the linearized regime, the system is generally underdetermined: multiple parameter displacements can yield (approximately) optimal empirical fit, making solutions across datasets incomparable due to non-identifiability. To obtain a canonical, comparable choice, we select the minimum-norm displacement among all (approximately) optimal solutions. This is equivalent to solving a regularized least-squares problem:
\begin{equation}
\Delta \theta :=\arg\min_{\Delta\theta} \left\|J(\mathcal{X};\theta) \Delta\theta - \big(\mathcal{Y} - f(\mathcal{X};\theta_0)\big)\right\|_2^2 + \lambda \left\|\Delta\theta\right\|_2^2.
\end{equation}
This objective is the parameter-space form of kernel ridge regression and admits the closed-form solution
\begin{equation}
\Delta \theta := \theta^*-\theta = J(\mathcal{X};\theta)^\top \mathbf \Theta^{-1} \left(\mathcal{Y} - f(\mathcal{X};\theta)\right),
\label{eq:parameter_displacement}
\end{equation}
where $\mathbf \Theta := J(\mathcal{X};\theta)J(\mathcal{X};\theta)^\top + \zeta I$ denotes the regularized NTK matrix. $\zeta = \frac{1}{\eta t}$ is a parameter corresponding to the early stopping model state at time $t$ with the learning rate $\eta$. We empirically set $\zeta=0.01$ in our implementations. Please refer to its sensitivity analysis in Section~\ref{sec:influence_factors}. Then we have the final value change caused by the projected displacement, 
\begin{equation}
V(\mathcal{D},\mathcal{T},\theta) \propto (\mathcal Y - f(\mathcal{X}_\mathcal{T};\theta))\langle \Delta\theta_\mathcal{D},\Delta\theta_\mathcal{T}\rangle/\|\Delta\theta_\mathcal{T}\|_2^2.
\label{eq:value_displacement_final}
\end{equation}

\subsection{MPC-friendly Optimization}
\label{sec:ntk_approximation}
To calculate the model parameter displacement, we mainly need the following computations: forward inference, Jacobian calculation, NTK matrix, and inversion.
Since the forward inference and Jacobian computation inherently offer limited room for efficiency improvement from the plaintext perspective, we focus our optimization efforts on approximating the NTK matrix and its associated computations. For NTK computation, one must evaluate pairwise gradient inner products across all samples, which requires $\frac{1}{2}N_\mathcal{D}^2$ inner products. Since these gradients are high-dimensional, precomputing and storing them in memory, GPU VRAM, or disk is infeasible, as MPC protocols keep them in secret-shared form, greatly inflating storage, communication, and computational costs.

To effectively improve efficiency, we propose a diagonal batch strategy to approximate the NTK matrix. Specifically, we partition the training dataset $\mathcal{D}$ into data batches $\{\mathcal{D}_{\mathrm{B}_i}=\{\bm{x}_{i, j}, \bm{y}_{i, j}\}_{j=1}^r\}_{i=1}^{N_\mathcal{D}/r}$ of size $r$, assuming that $N_\mathcal{D}$ is divisible by $r$. For each batch, we compute the gradients of all samples: $\{J(\bm{x}_{i, j}; \theta)\}_{j=1}^r$, and calculate the pairwise inner products between them, forming an $r \times r$ NTK batch block $\mathbf{\Theta}_{\mathrm{B}_i}$.
Each such block is then placed along the diagonal of a full NTK matrix $\widehat{\mathbf{\Theta}}(\mathcal X, \mathcal X;\theta)$, while all off-diagonal entries are set to zero.
This approximation yields two major benefits:
\begin{itemize}[leftmargin=*]
    \item \emph{Efficient Matrix Inversion}: Inverting $\widehat{\mathbf{\Theta}}(\mathcal X, \mathcal X;\theta)$ becomes equivalent to inverting each NTK batch block independently.
    \item \emph{Localized Computation}: When multiplying $J(\mathcal X; \theta)$ with $\widehat{\mathbf{\Theta}}(\mathcal X, \mathcal X;\theta)^{-1}$ that is needed by Eq.~\eqref{eq:parameter_displacement}, all cross-batch products are zero. Computations can be performed locally.
\end{itemize}

A key limitation of this approximation is that it neglects inter-batch correlations. One remedy is to repeatedly shuffle the data, compute a displacement for each permutation, and average the results~\cite{wu2022davinz}. However, this strategy is MPC-unfriendly: every shuffle requires a full displacement computation. Moreover, many permutations are needed to cover sufficient correlations. To address these issues, we introduce a landmark correlation strategy.

Specifically, we first uniformly sample a landmark set of size $Z$, $\mathcal{Z}=\{\bm{z}_i\}_{i=1}^{Z} \sim \mathbf{U}(\mathcal{D}, Z)$, then form $\mathbf{\Theta}_\mathrm{Z}=\mathbf{\Theta}(\mathcal{Z}, \mathcal{Z};\theta)\in\mathbb{R}^{Z\times Z}$ with entries $\Theta(\bm{z}_i,\bm{z}_{i^\prime};\theta)$; for each batch $\mathcal{D}_{\mathrm{B}_i}$, we form $\mathbf{\Theta}_{\mathrm{B}_i\mathrm{Z}}=\mathbf{\Theta}(\mathcal{X}_{\mathrm{B}_i},\mathcal{Z};\theta)\in\mathbb{R}^{r\times Z}$, $\mathbf{\Theta}_{\mathrm{B}_i}=\mathbf{\Theta}(\mathcal{X}_{\mathrm{B}_i},\mathcal{X}_{\mathrm{B}_i};\theta)\in\mathbb{R}^{r\times r}$, and the residual targets $\widetilde{\mathcal{Y}}_{\mathrm{B}_i}=\mathcal{Y}_{\mathrm{B}_i}-f(\mathcal{X}_{\mathrm{B}_i};\theta)$. We maintain a streaming global residual coefficient in the landmark span, parameterized by $\tau \in\mathbb{R}^{Z}$ such that the global component of the residual on batch $\mathcal{D}_{\mathrm{B}_i}$ is predicted by $\mathbf{\Theta}_{\mathrm{B}_i\mathrm{Z}}\tau$. Across batches, we accumulate only small $(Z\times Z)$ and $Z$-dimensional statistics about the NTK matrix and model outputs,
\begin{equation}
\Sigma_{\mathbf{\Theta}} \leftarrow \Sigma_{\mathbf{\Theta}} + \mathbf{\Theta}_{\mathrm{ZB}_i}\mathbf{\Theta}_{\mathrm{B}_i\mathrm{Z}},\qquad
\Sigma_{\mathcal{Y}} \leftarrow \Sigma_{\mathcal{Y}} + \mathbf{\Theta}_{\mathrm{ZB}_i}\widetilde{\mathcal{Y}}_{\mathrm{B}_i},
\label{eq:online_stats}
\end{equation}
Given these cumulative messages, we update the landmark coefficient by solving the linear system $\big(\Sigma_{\mathbf{\Theta}}+\mu\,\mathbf{\Theta}_{\mathrm{Z}}\big)\tau = \Sigma_{\mathcal{Y}}$,
\begin{equation}
\tau = \big(\Sigma_{\mathbf{\Theta}}+\mu\,\mathbf{\Theta}_{\mathrm{Z}}\big)^{-1}\, \Sigma_{\mathcal{Y}},
\label{eq:tau_closed_form}
\end{equation}
where $\mu>0$ controls the strength of the global regularization, and we empirically set $\mu=0.01$ in our implementation. Intuitively, $\mathbf{\Theta}_{\mathrm{ZB}_i}\mathbf{\Theta}_{\mathrm{B}_i\mathrm{Z}}$ measures how strongly the current batch constrains functions in the landmark span, while $\mathbf{\Theta}_{\mathrm{ZB}_i}\widetilde{\mathcal{Y}}_{\mathrm{B}_i}$ injects the corresponding residual evidence. For the current batch $\mathcal{D}_{\mathrm{B}_i}$, we first form the global prediction of its residual using the latest coefficient
$\tau$: $\widehat{\mathcal{Y}}_{\mathrm{B}_i} = \widetilde{\mathcal{Y}}_{\mathrm{B}_i} - \mathbf{\Theta}_{\mathrm{B}_i\mathrm{Z}}\tau$. We then compute a within-batch NTK update driven only by the corrected residual via a local damped solve: $\delta_{\mathrm{B}_i} \;=\; (\mathbf{\Theta}_{\mathrm{B}_i} + \zeta I_r)^{-1}\widehat{\mathcal{Y}}_{\mathrm{B}_i}$ and lift it back to parameter space through the batch Jacobian. After accumulating parameter displacements of all batches, we also need to add the displacement of the landmark:
\begin{equation}
\Delta\theta= J_Z^{\top}\tau_{N_\mathcal{D}/r} + \sum_{i=1}^{N_\mathcal{D}/r}J_{\mathrm{B}_i}^{\top}\delta_{\mathrm{B}_i},
\label{eq:online_param_update}
\end{equation}
where $\tau_{N_\mathcal{D}/r}$ is the landmark coefficient after the last batch calculation. In this way, we calculate the overall displacement with only $\mathcal O(Z^2)$ memory and a single pass over the batches.

\subsection{MPC Implementation of \method{}}
\label{sec:metho:mpc}
\begin{figure}[htbp]
\centering\includegraphics[width=0.99\linewidth]{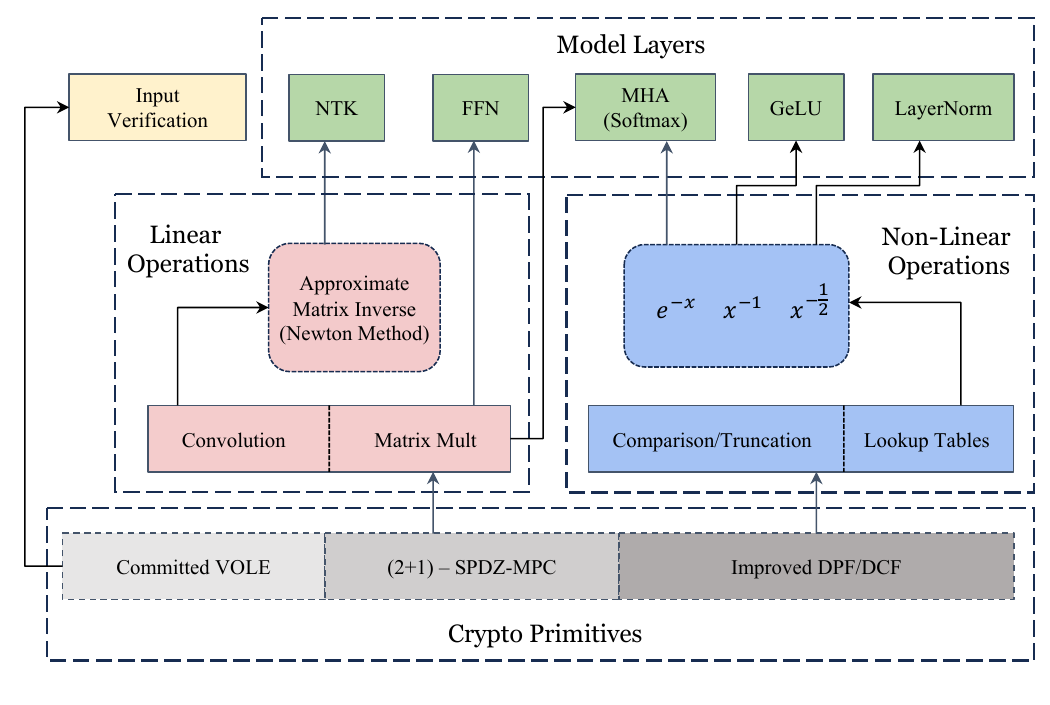}
\vspace{-25pt}
    \caption{System architecture (from the bottom to the top) for the MPC implementation of \method{}.}
    \label{fig:mpc:architecture}
    \vspace{-5pt}
\end{figure}

We focus on two research tasks: (1) input verification with respect to the seller's commitment to the dataset, and (2) the private evaluation of NTK, multi-head attention, feedforward, and other functions that appear in \method{}. 
An overview of our MPC framework can be found in Figure~\ref{fig:mpc:architecture}. 

\shortsection{Input Verification.}
As introduced in Section~\ref{sec:prelim:mpc}, we assume that the commitment to dataset $\mathcal D$, denoted as ${\sf com}_\mathcal{D}$, is known publicly. Upon secret-sharing $\rshare{\mathcal D}$ in MPC, $P_0$ is required to prove that ${\sf com}_\mathcal{D}$ and $\rshare{\mathcal D}$ are consistent. It ensures that the same dataset is used across the data trading process.

\paragraph{Problem statement.}
The input verification can naively be done by computing the decommitment function ${\sf Open}({\sf com}_\mathcal{D}, [{\sf decom}_\mathcal{D}])$ privately inside MPC, in which ${\sf decom}_\mathcal{D}$ usually contains $[\mathcal{D}]$ and optional randomness. However, canonical commitment schemes either compute over exponents (e.g., Pedersen~\cite{pedersen1991non}) or rely on collision-resistant hash functions (e.g., Merkle tree~\cite{merkle1987digital}), which are time-consuming when computed in MPC.

\paragraph{Our result.}
We incorporate the committed vector oblivious linear evaluation (C-VOLE)~\cite{sun2025committed} with the maliciously secure SPDZ-MPC protocol.
Our design lets parties prove the validity of their inputs via C-VOLE, then convert the output of C-VOLE to authenticated secret shares in MPC for free. Previous work builds a Merkle tree from SHA-256, and it takes $529$ seconds\footnote{We do not take the LowMC-128 and AES-128 implemented in~\cite{peng2025reliable} because they do not yield a secure collision-resistant hash function when instantiated in Davies-Meyer construction (due to the birthday bound). } to evaluate $\ell=2^{20}$ bits of data~\cite{peng2025reliable}. Concretely, a SHA-256 circuit is of size $n_c = 22573$ and depth $n_d = 1607$. It requires extra communication overhead and round trips linear to $n_c$ and $n_d$, respectively. 
Our approach only needs $\mathcal O({\sf polylog} \ell)$ communication and $2$ rounds.

\paragraph{Protocol sketch.}
Assume that $P_0$ commits to $\vecx$. We first describe C-VOLE, then describe the conversion from C-VOLE to $\rshare{\vecx}$.
C-VOLE relies on a special hybrid commitment defined as ${\sf com}_{\vecx} = ({\sf com}_{\vecr}, \vecd, H)$, in which ${\sf com}_{\vecr}$ is a commitment to a sparse random vector $\vecr$, $\vecd$ is defined as $\vecd = \vecx - \vecr \cdot H$, and $H$ is a public random matrix. It satisfies both binding (i.e., cannot be opened to another value $\vecx'\neq \vecx$) and hiding (i.e., does not reveal information about $\vecx$). What makes it useful to our scheme is that during the proof of opening phase, $P_0$ can leverage C-VOLE protocols to generate a VOLE correlation on $\vecr$ while proving the consistency of $\vecr$ and ${\sf com}_{\vecr}$. At the end of C-VOLE, it lets $P_1$ authenticate $\vecr$ by obtaining a local key $K_{\vecr}$. $P_0$ holds a MAC $M_{\vecr}$ s.t. $M_{\vecr} = K_{\vecr} + \vecr \cdot \Delta_1$. Our observation is that they can locally convert the authenticated $\vecr$ directly to an authenticated secret share $[\vecx]$. To achieve this, $P_0$ defines $M_0^{\vecx} = \vecr \cdot H \cdot \Delta_0 + M_{\vecr} \cdot H + \vecd \cdot \Delta_0$ and $P_1$ defines $M_1^{\vecx} = -K_{\vecr} \cdot H + \vecd \cdot \Delta_1$. Recall that $\vecd = \vecx - \vecr \cdot H$, we have
$$M_0^{\vecx} + M_1^{\vecx} = (\vecr \cdot H + \vecd) \cdot (\Delta_0 + \Delta_1) = \vecx \cdot (\Delta_0 + \Delta_1),$$
which is a valid SPDZ authenticated secret share $[\vecx]$ as described in Section~\ref{sec:prelim:mpc}. Regarding the overhead, the transmission of $\vecd$ can be viewed as part of the share distribution process. C-VOLE for $\vecr$ only incurs $\mathcal O({\sf polylog}(|\vecx|))$ and all other procedures are executed locally. The algorithm description is in Supplementary Material~\cite{secdva_sm}.

\smallskip \shortsection{Maliciously Secure (2+1)-PC for DVA.}
After the input verification, the DVA will be conducted on the authenticated secret-shared dataset. 
This process utilizes the cryptographic secure computing techniques to ensure that no information is revealed beyond the output of DVA.
To achieve this goal, we build our (2+1)-party maliciously secure MPC framework to efficiently realize \method{}. It consists of a semi-honest, non-colluding dealer $P^*$ and two parties $P_0$ and $P_1$, who may be subject to malicious corruption.
In practice, the data trading platform takes the role of the dealer, and the data seller and buyer act as $P_0$ and $P_1$.

\paragraph{Problem statement.}
Building \method{} poses two technical challenges regarding accuracy, security, and efficiency.
\begin{itemize}[leftmargin=*]
    \item Existing MPC implementations of GeLU, Softmax, and LayerNorm are optimized for inference with minor precision requirements. In those schemes, these functions are highly approximated, and numerical errors are accumulated across layers and eventually degrade the precision. Such loss is only acceptable for inference as it only needs approximate correctness (e.g., top-1 label), not exact numerical fidelity. However, in \method{}, the computations such as gradients and matrix inversion are extremely precision-sensitive. It is non-trivial to adapt existing approaches to high-precision evaluations without sacrificing performance.
    \item Achieving malicious security when computing nonlinear functions in MPC is inherently challenging and often incurs substantially higher overhead than semi-honest protocols. Although private numerical computation based on primitives such as DPF and DCF are highly efficient, it's unclear how to adapt them to malicious settings. It leaves open questions of how to efficiently realize \method{} under the $(2+1)$-party maliciously secure setting.
\end{itemize}

\paragraph{Our results.} 
An overview of our MPC architecture can be found in Figure~\ref{fig:mpc:architecture}.
To efficiently and accurately emulate the math functions (i.e., GeLU, Softmax, LayerNorm on the right-hand side), we compose the piecewise polynomial approximation method with the oblivious lookup tables. Both approaches turn nonlinear function evaluations into MPC-friendly arithmetic and range checks, and will eventually be realized by integrating DPF/DCF into our SPDZ $(2+1)$-PC. Additionally, we explore the Newton method for approximate matrix inverse computation (i.e., in the middle) and propose a dynamic precision trick to maintain its accuracy.

To achieve \emph{malicious security}, we design lightweight consistency checks that validate the above arithmetic, range checks, PIR, lookup table, and dynamic precision for fixed-point representation of real numbers. Especially, we incorporate the DPF into the SPDZ $(2+1)$-PC and build a logarithmic-cost maliciously secure lookup table. Further optimization extends it to efficient comparison and range checks.
We present our improvement on maliciously secure DPF/DCF, as they are the most prominent contributions. Our complete MPC framework and security proof are in Supplementary Material~\cite{secdva_sm}.

\begin{table}[t]
  \centering
  \footnotesize
    \caption{Communication (per party) and computation overhead for distributed point and comparison functions. Table includes schemes from Half-tree (HT)~\cite{guo2023half}, Shark~\cite{gupta2025shark}, and Grotto~\cite{storrier2023grotto}. Some small terms are ignored. $n,\ell,\lambda,\kappa$ are input bit size, logarithmic table size and security parameters. DPF assumes full evaluation to implement table lookup.}
    \vspace{-5pt}
    \setlength{\tabcolsep}{2pt}
  \begin{tabular}{cccc}
    \toprule
    \multirow{2}{*}{Scheme} & \multirow{2}{*}{Communication} & \multicolumn{2}{c}{Computation} \\
    & & Gen & Eval \\\midrule
    HT DPF (semi-honest) & $\ell\kappa$ & $2\ell$ & $2^\ell$ \\
    HT DCF (semi-honest)  & $n\kappa + n^2$ & $4n$ & $2n$ \\\midrule
    Shark DPF & $\ell\kappa$ & $4\ell$ & $2^{\ell+1}$ \\
    Shark DCF & $n\kappa+n^2$ & $8n$ & $2n$ \\\midrule
    HT DPF + SPDZ (Our DPF) & $\ell\kappa$  & $2\ell$ & $2^\ell$ \\
    HT DPF + Grotto DCF + SPDZ (Our DCF) & $n\kappa$ & $2n$ & $n$ \\
    \bottomrule
  \end{tabular}
  \label{tab:dpf:analysis}
  \vspace{-10pt}
\end{table}

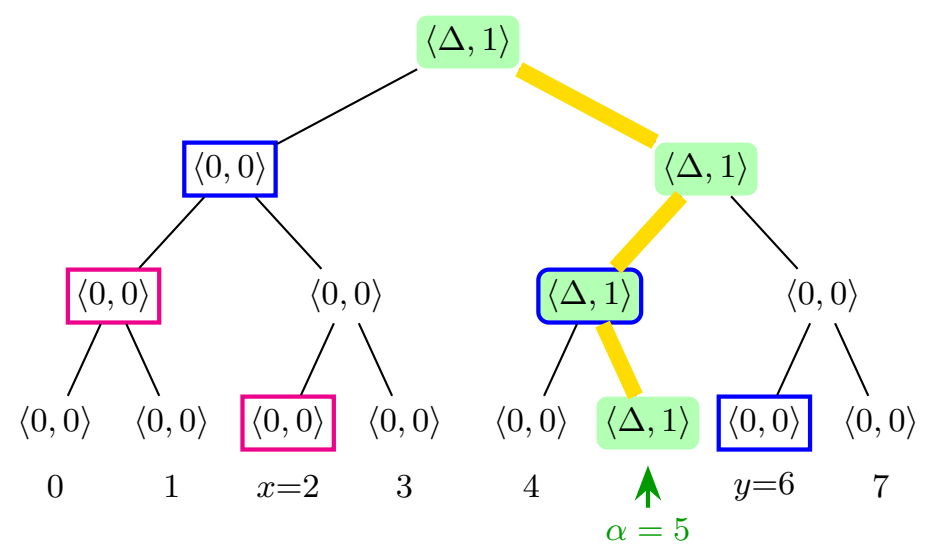
\begin{figure}[htbp]
\centering
\begin{tikzpicture}[
    scale=0.7,
    transform shape,
    level 1/.style={sibling distance=45mm, level distance=12mm},
    level 2/.style={sibling distance=22mm, level distance=12mm},
    level 3/.style={sibling distance=11mm, level distance=12mm},
    every node/.style={minimum width=0.8cm, minimum height=0.5cm, inner sep=2pt, font=\small},
    greennode/.style={fill=green!30, rounded corners=2pt},
    blueframe/.style={draw=blue, thick, rectangle},
    pinkframe/.style={draw=magenta, thick, rectangle},
    yellowedge/.style={draw=yellow!80!orange, line width=3pt},
    redlabel/.style={red, font=\normalsize},
]
\node[greennode] (root) {$\langle \Delta, 1 \rangle$}
    child {node[blueframe] (l1) {$\langle 0,0 \rangle$}
        child {node[pinkframe] (l2l) {$\langle 0,0 \rangle$}
            child {node (l3ll) {$\langle 0,0 \rangle$}}
            child {node (l3lr) {$\langle 0,0 \rangle$}}
        }
        child {node (l2r) {$\langle 0,0 \rangle$}
            child {node[pinkframe] (l3rl) {$\langle 0,0 \rangle$}}
            child {node (l3rr) {$\langle 0,0 \rangle$}}
        }
    }
    child {node[greennode] (r1) {$\langle \Delta,1 \rangle$}
        child {node[greennode, blueframe] (r2l) {$\langle \Delta,1 \rangle$}
            child {node (r3ll) {$\langle 0,0 \rangle$}}
            child {node[greennode] (r3lr) {$\langle \Delta,1 \rangle$}}
        }
        child {node (r2r) {$\langle 0,0 \rangle$}
            child {node[blueframe] (r3rl) {$\langle 0,0 \rangle$}}
            child {node (r3rr) {$\langle 0,0 \rangle$}}
        }
    };

\draw[yellowedge] (root) -- (r1);
\draw[yellowedge] (r1) -- (r2l);
\draw[yellowedge] (r2l) -- (r3lr);

\draw[-{Stealth}, green!60!black, thick] (r3lr) ++(0, -0.8) -- ++(0, 0.4);
\node[green!60!black] at ([yshift=-1.0cm]r3lr) {$\alpha = 5$};

\node at ([yshift=-0.6cm]l3ll) {$0$};
\node at ([yshift=-0.6cm]l3lr) {$1$};
\node at ([yshift=-0.6cm]l3rl) {$x{=}2$};
\node at ([yshift=-0.6cm]l3rr) {$3$};
\node at ([yshift=-0.6cm]r3ll) {$4$};
\node at ([yshift=-0.6cm]r3rl) {$y{=}6$};
\node at ([yshift=-0.6cm]r3rr) {$7$};

\end{tikzpicture}
\vspace{-5pt}
\caption{An example for the evaluation of DCF. To compute the output for $x=2$, we sum the blocks in {\color{purple}purple} and get $\ang{0,0}$, which means $x<\alpha$. For $y=6$, we sum the blocks in {\color{blue}blue} and get $\ang{\Delta,1}$, which means $y\ge\alpha$. }
\label{fig:alpha-beta-tree}
\end{figure}

\paragraph{Protocol sketch.} 
DPF/DCF are the most impactful techniques for nonlinear function evaluation in the $(2+1)$-PC setting. Taking Figure~\ref{fig:alpha-beta-tree} as an example, DPF allows $P^*$ to encode a secret value $\alpha=5$ into a binary tree structure and secretly share it with two parties. Each node on the tree encodes a value, either `0' or `1'. $P_0$ and $P_1$ hold a value $x$ and will evaluate to $x$-th path. Eventually, they obtain a \emph{secret share} of $1$ only if $x = \alpha$ (by traversing the highlighted path), otherwise they obtain a $0$. DCF requires applying a more complicated encoding, which lets parties obtain $1$ only if $x \geq \alpha$ (by traversing to any path indexed by $i\geq 5$). Our first contribution utilizes the \emph{half-tree encoding}~\cite{guo2023half} to encode the IT-MAC with the secret-shared values, which allows the parties to obtain \emph{authenticated secret shares} after the path evaluation. This design incorporates DPF/DCF with SPDZ maliciously secure MPC and allows secure computing of linear arithmetic, PIR, lookup table, secure comparison, and range checks. Additionally, we apply Grotto's optimization to half-tree, which builds a DCF directly from its DPF, instead of using the more complicated encoding. Taking Figure~\ref{fig:alpha-beta-tree} as an example, evaluating $x = 2$ only requires reaching two nodes labeled purple. The sum of labels on these nodes indicates that none of the $\{0,1,2\}$-th leaf node is $1$, hence $x \geq \alpha$ is false. In another case of $x = 6$, evaluating $3$ nodes labeled in blue is sufficient to deduce that $x \geq \alpha$, since one of them captures the $1$ label. Overall, for $n$-bit inputs, two parties can obtain the result of $x \geq \alpha$ by simply evaluating at most $n$ nodes and summing up the attached labels.

The comparison in Table~\ref{tab:dpf:analysis} shows that the cost of our DPF aligns with the semi-honest protocol and is $2\times$ more lightweight than the state-of-the-art Shark~\cite{gupta2025shark}. Our DCF gets rid of the $\mathcal O(n^2)$ factor from others and also requires less computation than previous works.
\begin{table}[h]
\centering
\caption{Performance comparison between \method{} and other baseline approaches when assessing the value of CIFAR10/CIFAR100 to enhance the ViT-Tiny performance on CIFAR100/CIFAR10 in plain-text settings.}
\vspace{-5pt}
\resizebox{.45\textwidth}{!}{
\setlength{\tabcolsep}{5pt}
\hspace{-5pt}
\begin{tabular}{lccc|ccc}
\toprule
 & \multicolumn{3}{c|}{\textbf{CIFAR10 $\rightarrow$ CIFAR100}} & \multicolumn{3}{c}{\textbf{CIFAR100 $\rightarrow$ CIFAR10}}\\
\textbf{Method} & P.C.$\uparrow$ & S.C.$\uparrow$ & Cost (s) & P.C.$\uparrow$ & S.C.$\uparrow$ & Cost (s)\\
\midrule
DAVINZ    &-0.632&-0.400&3923.4&-0.921&-0.800&3950.7  \\
FreeShap    &0.394&0.300&7698.3&-0.498&-0.600&7862.2  \\
In-Run DS    &0.529&0.400&1108.8&0.445&-0.200&1094.5  \\
WTE    &0.500&0.100&2980.5&0.804&0.800&2907.6  \\
TV    &0.825&0.900&3098.9&0.826&0.800&3091.4  \\ \midrule
\method{}    &\textbf{0.843}&\textbf{0.900}&\textbf{207.4}&\textbf{0.855}&\textbf{0.900}&\textbf{192.5} \\
\bottomrule
\end{tabular}}
\label{tab:exp_cifar10/100}
\end{table}
\begin{table*}[h]
\centering
\caption{Performance comparison between \method{} and other baselines when assessing the value of different domains in PACS and Office-Home to enhance the ViT-Tiny performance on leaving one domain out as the target in plain-text settings.}
\vspace{-5pt}
\resizebox{.99\textwidth}{!}{
\setlength{\tabcolsep}{5pt}
\hspace{-5pt}
\begin{tabular}{lcc cc cc cc c|cc cc cc cc c}
\toprule
& \multicolumn{9}{c|}{\textbf{PACS (with domain P, A, C, S)}} & \multicolumn{9}{c}{\textbf{Office-Home (domain A, C, P, R)}} \\
 & \multicolumn{2}{c}{\textbf{A, C, S $\rightarrow$ P}} & \multicolumn{2}{c}{\textbf{P, C, S $\rightarrow$ A}} & \multicolumn{2}{c}{\textbf{P, A, S $\rightarrow$ C}} & \multicolumn{2}{c}{\textbf{P, A, C $\rightarrow$ S}} & Avg. & \multicolumn{2}{c}{\textbf{C, P, R $\rightarrow$ A}} & \multicolumn{2}{c}{\textbf{A, P, R $\rightarrow$ C}} & \multicolumn{2}{c}{\textbf{A, C, R $\rightarrow$ P}} & \multicolumn{2}{c}{\textbf{A, C, P $\rightarrow$ R}} & Avg.\\
\textbf{Method} & P.C.$\uparrow$ & S.C.$\uparrow$  & P.C.$\uparrow$ & S.C.$\uparrow$& P.C.$\uparrow$ & S.C.$\uparrow$& P.C.$\uparrow$ & S.C.$\uparrow$ & Cost (s) & P.C.$\uparrow$ & S.C.$\uparrow$& P.C.$\uparrow$ & S.C.$\uparrow$& P.C.$\uparrow$ & S.C.$\uparrow$& P.C.$\uparrow$ & S.C.$\uparrow$ & Cost (s) \\
\midrule
DAVINZ    &0.668&0.600&0.435&0.486&0.771&0.543&-0.302&-0.086&798.5&0.486&0.314&0.703&0.371&0.546&0.600&\textbf{0.844}&0.714&1276.9 \\
FreeShap    &0.725&0.714&0.081&0.200&0.218&-0.086&-0.428&-0.200&6030.4&0.032&0.029&-0.221&-0.143&-0.108&0.257&0.358&0.200&6578.0 \\
In-Run DS    &0.501&0.429&0.750&0.829&-0.607&-0.543&0.878&\textbf{0.943}&387.3&0.328&0.543&-0.571&-0.543&-0.246&0.143&0.613&0.714&562.9 \\
WTE    &0.755&0.657&\textbf{0.969}&0.870&-0.811&-0.714&0.781&0.771&125.6&0.220&0.429&\textbf{0.985}&0.943&\textbf{0.927}&0.371&-0.034&-0.486&172.7 \\
TV    &0.681&0.771&0.825&0.771&-0.284&-0.200&0.710&0.486&1728.8&0.396&0.486&0.414&0.257&-0.516&-0.486&0.111&-0.429&2342.2 \\ \midrule
\method{}   &\textbf{0.862}&\textbf{0.829}&0.928&\textbf{0.886}&\textbf{0.922}&\textbf{0.943}&\textbf{0.914}&0.886&\textbf{89.2}&\textbf{0.833}&\textbf{0.829}&0.971&\textbf{0.943}&0.894&\textbf{0.886}&0.835&\textbf{0.829}&\textbf{147.6} \\
\bottomrule
\end{tabular}}
\label{tab:exp_pacs/officehome}
\end{table*}
\begin{table*}[h]
\centering
\caption{Performance comparison between \method{} and other baseline approaches when assessing the value of different domains in DomainNet to enhance the ViT-Tiny performance on every domain of PACS and Office-Home in plain-text settings.}
\vspace{-5pt}
\resizebox{.99\textwidth}{!}{
\setlength{\tabcolsep}{5pt}
\hspace{-5pt}
\begin{tabular}{lcc cc cc cc c|cc cc cc cc c}
\toprule
& \multicolumn{9}{c|}{\textbf{DomainNet$\rightarrow$PACS (with domain P, A, C, S)}} & \multicolumn{9}{c}{\textbf{DomainNet$\rightarrow$Office-Home (with domain A, C, P, R)}} \\
 & \multicolumn{2}{c}{\textbf{P}} & \multicolumn{2}{c}{\textbf{A}} & \multicolumn{2}{c}{\textbf{C}} & \multicolumn{2}{c}{\textbf{S}} & Avg. & \multicolumn{2}{c}{\textbf{A}} & \multicolumn{2}{c}{\textbf{C}} & \multicolumn{2}{c}{\textbf{P}} & \multicolumn{2}{c}{\textbf{R}} & Avg.\\
\textbf{Method} & P.C.$\uparrow$ & S.C.$\uparrow$  & P.C.$\uparrow$ & S.C.$\uparrow$& P.C.$\uparrow$ & S.C.$\uparrow$& P.C.$\uparrow$ & S.C.$\uparrow$ & Cost (s) & P.C.$\uparrow$ & S.C.$\uparrow$& P.C.$\uparrow$ & S.C.$\uparrow$& P.C.$\uparrow$ & S.C.$\uparrow$& P.C.$\uparrow$ & S.C.$\uparrow$ & Cost (s) \\
\midrule
DAVINZ    &0.908&0.886&0.869&0.943&0.036&0.486&0.048&-0.086&3723.8&0.711&0.829&0.384&0.600&0.470&0.371&0.802&0.771&3894.2 \\
FreeShap    &-0.526&-0.600&0.378&0.657&-0.242&-0.143&-0.900&-0.829&7279.4&0.891&0.943&-0.746&-0.771&-0.201&-0.543&-0.595&-0.257&7869.2 \\
In-Run DS    &0.288&0.029&0.889&0.771&0.810&0.829&0.824&0.886&923.3&0.883&0.886&0.766&0.600&0.673&0.543&\textbf{0.902}&0.943&1447.6 \\
WTE    &0.444&0.200&0.443&0.257&-0.373&-0.886&0.748&0.771&2679.0&0.855&0.886&0.447&0.314&-0.006&0.429&0.594&0.667&2862.4 \\
TV    &0.923&0.771&\textbf{0.952}&0.943&0.709&0.600&0.779&0.543&3408.3&\textbf{0.934}&0.600&0.624&0.714&0.891&0.371&0.775&0.771&3124.1 \\ \midrule
\method{}   &\textbf{0.934}&\textbf{0.886}&0.945&\textbf{0.943}&\textbf{0.825}&\textbf{0.829}&\textbf{0.836}&\textbf{0.886}&\textbf{245.9}&0.859&\textbf{0.943}&\textbf{0.835}&\textbf{0.829}&\textbf{0.940}&\textbf{0.829}&0.878&\textbf{0.943}&\textbf{280.2} \\
\bottomrule
\end{tabular}}
\label{tab:exp_domainnet}
\end{table*}



\section{Experiments}
\noindent \textbf{Datasets.}
We use CIFAR-10~\cite{cifar} and CIFAR-100~\cite{cifar} to simulate scenarios with only label discrepancy. To study feature discrepancy, we adopt PACS~\cite{pacs} and Office-Home~\cite{office-home}, each of which consists of 4 data domains. For cases involving both feature and label discrepancies, we adopt the largest available cross-domain dataset, DomainNet~\cite{domainnet}, and experiment with it on transferring to PACS and Office-Home. 

\smallskip \noindent \textbf{Implementation Details.}
We adopt ViT-Tiny~\cite{vit} as the target model.
To improve efficiency, we only conduct computations on the last attention layer of the target model. The task module for different datasets is a single linear layer. For NTK matrix approximation, we set the batch size to 32, the landmark set size to 1\% of the dataset size, and use $\zeta = 0.01$ and $\mu=0.01$. 
For the MPC implementation, we use the ring $\mathbb{Z}_{2^{128}}$ to represent the secret sharing and IT-MACs. The values and global keys are represented as integers in $\mathbb{Z}_{2^{64}}$. We approximate floating-point numbers using 12-bit fixed-point representation. The implementation is primarily built upon \texttt{emp-tools}~\cite{emp-toolkit}. We evaluate the MPC implementations on a server equipped with an AMD EPYC 9555 processor and 512 GB of RAM. All experiments are executed in a single thread, and the communication between parties is simulated over a local network environment with a bandwidth of 10 Gbps and 0.01 ms RTT, similar to FullPto~\cite{peng2025reliable}.

\smallskip \noindent \textbf{Comparison Baselines.}
We compare \method{} with state-of-the-art methods in data valuation (DV) and data selection (DS). For DV, we adopt the most lightweight training-free baselines: DAVINZ~\cite{wu2022davinz}, In-Run DS~\cite{in-runDV}, and FreeShap~\cite{freeshap}. 
For DS methods, we benchmark against WTE~\cite{liu2025wasserstein} and Task Vector (TV)~\cite{tv}. 
For MPC implementation, we first compare \method{} with FullPto's input verification~\cite{peng2025reliable}. Then, we primarily compare with Shark~\cite{gupta2025shark} in executing major computations of \method{}. Since MPC performance varies across settings and our protocol targets the $(2+1)$-party setup, we exclude comparisons with other configurations. Shark~\cite{gupta2025shark}, following the same setup, serves as our state-of-the-art baseline. Besides, we only compare our enhanced $(2+1)$-party protocol with Shark in inference, as \method{} involves computations unsupported by Shark.

\smallskip \noindent \textbf{Evaluation Metrics.}
To measure the ground-truth data value, we adapt the task transfer distance~\cite{achille2019task2vec} with details provided in Supplementary Material~\cite{secdva_sm}.  
We evaluate different approaches by computing their Pearson and Spearman correlations (P.C. and S.C., respectively, the higher the better) with this ground-truth value. We assess the efficiency by measuring the wall-clock computational cost. For MPC experiments, we do not distinguish between online and offline operations; instead, we calculate the total overall overhead.

\subsection{Effectiveness of \method{} in Plain-text Setup}
\subsubsection{\textbf{Comparison to baseline approaches.}}
To simulate the \textit{label-discrepancy-only} scenarios, we partition CIFAR-10 and CIFAR-100 by class and create five splits to be assessed. Each split contains 10,000 randomly selected samples drawn from an increasing proportion of classes, with a stride of 20\% of the total classes. The buyer’s target-task testing set consists of 2,000 randomly sampled examples from the full class set. For \textit{feature-discrepancy-only} scenarios, we adopt a leave-one-domain-out protocol on PACS and Office-Home. In each case, the held-out domain serves as the buyer's testing set, while each remaining domain provides randomly selected samples (repeated twice) as the seller's dataset. The testing set size is fixed to half the size of the selling set: 1,600 samples for PACS and 2,400 for Office-Home. For the combined \textit{feature-label-discrepancy} scenarios, we randomly select 5,000 samples from each DomainNet domain as the seller's datasets, while 800 and 1,200 from each domain of PACS and Office-Home, respectively, are used as the buyer's test sets. According to experiment results shown in Tables~\ref{tab:exp_cifar10/100}, \ref{tab:exp_pacs/officehome}, and \ref{tab:exp_domainnet}, we can observe that \textit{\textbf{\method{} can mostly provide the most accurate data value estimation across dataset pairs in the shortest time.}} In contrast, baseline approaches cannot perform stably well. For example, training-based approaches, such as In-Run DS and TV, generally perform better than training-free ones, but they perform poorly in assessing the value of Office-Home with feature discrepancies.
For other DV approaches, DAVINZ generally performs better than FreeShap. 
As for DS methods, TV performs quite well in scenarios with feature-label discrepancy, but such good performance does not hold in other scenarios.

\begin{table}[h]
\centering
\caption{Ablation study of different designs of \method{} in plain-text settings. The experiments assess the value of different domains in DomainNet in enhancing the ViT-Tiny performance on every domain of Office-Home.}
\vspace{-5pt}
\resizebox{.48\textwidth}{!}{
\setlength{\tabcolsep}{3pt}
\hspace{-5pt}
\begin{tabular}{lcc cc cc cc c}
\toprule
& \multicolumn{9}{c}{\textbf{DomainNet$\rightarrow$OfficeHome (with domain A, C, P, R)}} \\
 & \multicolumn{2}{c}{\textbf{A}} & \multicolumn{2}{c}{\textbf{C}} & \multicolumn{2}{c}{\textbf{P}} & \multicolumn{2}{c}{\textbf{R}} & Avg.\\
\textbf{Method} & P.C.$\uparrow$ & S.C.$\uparrow$  & P.C.$\uparrow$ & S.C.$\uparrow$& P.C.$\uparrow$ & S.C.$\uparrow$& P.C.$\uparrow$ & S.C.$\uparrow$ & Cost (s) \\
\midrule
\method{} w/o L. Cor.   &0.745&0.714&0.675&0.543&0.802&0.543&0.745&0.829&215.7 \\
\method{} w/o C. Cor.   &0.817&0.829&0.803&0.829&0.864&0.714&0.789&0.829&261.6 \\ \midrule
\method{}   &0.859&0.943&0.835&0.829&0.940&0.829&0.878&0.943&280.2 \\
\bottomrule
\end{tabular}}
\label{tab:ablation}
\end{table}
\begin{table}[h]
\centering
\caption{Sensitivity analysis of model scales in plain-text settings. The experiments assess the value of different domains in DomainNet to Office-Home.}
\vspace{-5pt}
\resizebox{.48\textwidth}{!}{
\setlength{\tabcolsep}{3pt}
\hspace{-5pt}
\begin{tabular}{lcc cc cc cc c}
\toprule
& \multicolumn{9}{c}{\textbf{DomainNet$\rightarrow$OfficeHome (with domain A, C, P, R)}} \\
 & \multicolumn{2}{c}{\textbf{A}} & \multicolumn{2}{c}{\textbf{C}} & \multicolumn{2}{c}{\textbf{P}} & \multicolumn{2}{c}{\textbf{R}} & Avg.\\
\textbf{Method} & P.C.$\uparrow$ & S.C.$\uparrow$  & P.C.$\uparrow$ & S.C.$\uparrow$& P.C.$\uparrow$ & S.C.$\uparrow$& P.C.$\uparrow$ & S.C.$\uparrow$ & Cost (s) \\
\midrule
\method{} on ViT-Large   &0.855&0.943&0.848&0.886&0.910&0.829&0.862&0.886&9377.4 \\
\method{} on ViT-Base   &0.846&0.886&0.840&0.829&0.931&0.829&0.878&0.943&3120.6 \\ 
\method{} on ViT-Small   &0.834&0.886&0.848&0.886&0.912&0.829&0.885&0.943&829.8 \\ \midrule
\method{} on ViT-Tiny   &0.859&0.943&0.835&0.829&0.940&0.829&0.878&0.943&280.2 \\
\bottomrule
\end{tabular}}
\label{tab:model_scale}
\end{table}

\subsubsection{\textbf{Ablation study and sensitivity analysis.}}
\label{sec:influence_factors}
We conduct a comprehensive ablation study in the DomainNet to Office-Home cases, as shown in Table~\ref{tab:ablation}. We first detach the use of the landmark and only use the diagonal batch NTK approximation. Then, we validate the cumulative correction offered by each batch, i.e., the computation of $\Sigma_\mathbf{\Theta}$ and $\Sigma_\mathcal{Y}$ within each batch as $\mathbf{\Theta}_{\mathrm{ZB}_i}\mathbf{\Theta}_{\mathrm{B}_i\mathrm{Z}}$ and $\mathbf{\Theta}_{\mathrm{ZB}_i}\widetilde{\mathcal{Y}}_{\mathrm{B}_i}$, respectively. By comparing \method{} with `\method{} w/o L. Cor.' and `\method{} w/o C. Cor.', the necessity of landmark correction and cumulative correction can be validated.

\begin{figure}[htbp]
  \centering
  \begin{subfigure}{0.48\linewidth}
    \centering
\includegraphics[width=\linewidth]{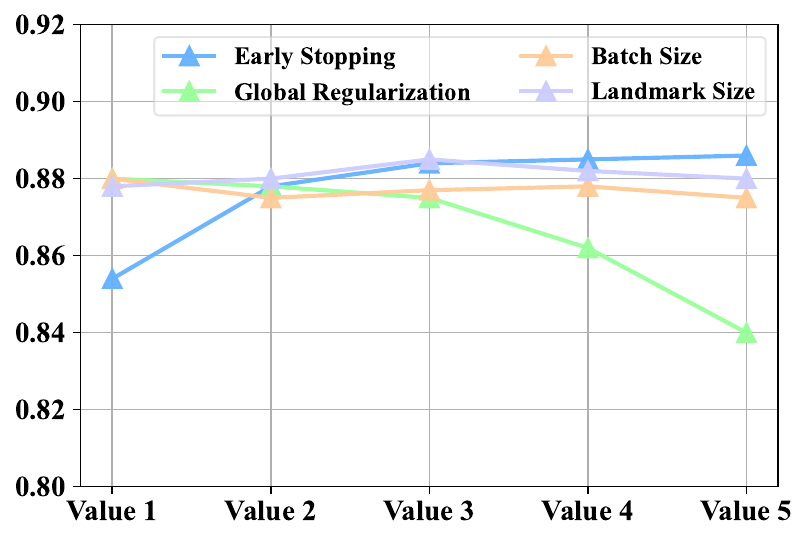}
    \vspace{-10pt}
    \caption{Average Pearson Correlation}
    \label{fig:left}
  \end{subfigure}
  \,
  \begin{subfigure}{0.48\linewidth}
    \centering
    \includegraphics[width=\linewidth]{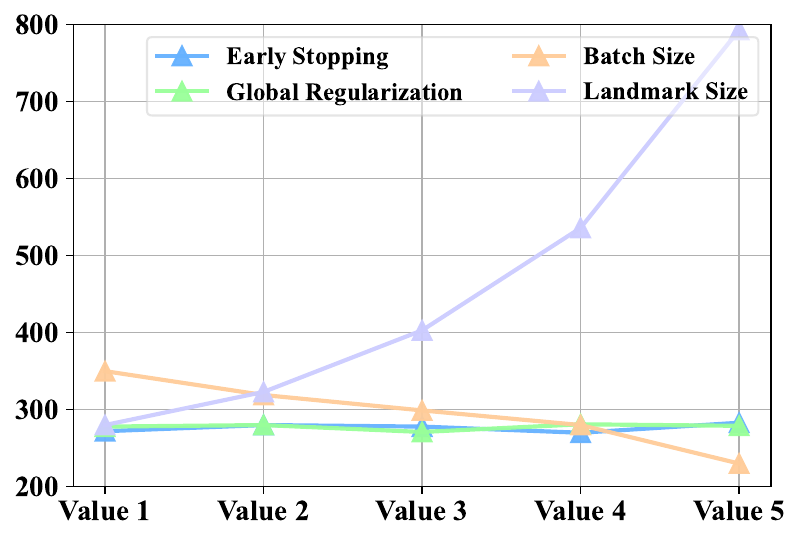}
    \vspace{-10pt}
    \caption{Average Cost (s)}
    \label{fig:right}
  \end{subfigure}
  \vspace{-10pt}
  \caption{Sensitivity analysis on early stopping parameter $\zeta$, global regularization coefficient $\mu$, batch size, and landmark size at five different values. The experiments are conducted on transferring DomainNet to Office-Home.}
  \label{fig:pair}
\end{figure}

We conduct a sensitivity analysis of the model scale and hyperparameters of \method{} in the DomainNet to Office-Home cases. Specifically, in addition to ViT-Tiny, we conduct experiments on ViT-Small, ViT-Base, and ViT-Large. The results shown in Table~\ref{tab:model_scale} demonstrate the effective scalability of \method{}. Then, we employ a series of early stopping parameters ($\zeta$=0.1, 0.01, 0.001, 0.0001, 0.00001), global regularization coefficient ($\mu$=0.1, 0.01, 0.001, 0.0001, 0.00001), batch sizes (4, 8, 16, 32, 64), and landmark sizes (1\%, 2\%, 5\%, 10\%, 20\%), and the experiment results are shown in Figures~\ref{fig:pair}. We can observe that when the early stopping parameter is lower than a threshold (approximately between 0.01 and 0.001), the performance of \method{} becomes stable, which may correspond to the NTK convergence state. As for $\mu$, the performance is stable when it is larger than 0.001, but drops when it becomes too small. Besides, the performance of \method{} is not sensitive to the batch size, but the computation time increases when the batch size decreases. An inverse phenomenon can also be observed for the landmark size. 

\begin{table}[h]
  \centering
  \footnotesize
    \caption{Comparison between \method{} and FullPto~\cite{peng2025reliable} in the input verification for a $32$MB dataset.}
    \vspace{-5pt}
  \begin{tabular}{cccc}
    \toprule
    \multirow{2}{*}{Schemes} & \multirow{2}{*}{Time} & \multicolumn{2}{c}{Security Parameter} \\
                            & & Computational & Statistical \\\midrule
    FullPto LowMC & $3.49$ s & 64 & $0.01$ \\
    FullPto SHA-256 & $119$ s & 128 & $0.01$ \\
    Ours & $2.09$ ms & 128 & $2^{-40}$ \\
    \bottomrule
  \end{tabular}
  \label{tab:input:verification}
  \vspace{-10pt}
\end{table}

\subsection{Effectiveness of \method{} in MPC Setup}
\label{sec:perf:mpc}

\subsubsection{\textbf{Input Verification.}}
Table~\ref{tab:input:verification} compares the C-VOLE input verification~\cite{sun2025committed} with the hash-based verification from~\cite{peng2025reliable}. 
It shows that our input verification is $1670\times$ faster than its less-secure LowMC version and $56938\times$ faster than its secure SHA-256 version. The reason is that our scheme employs ZKP to avoid MPC computation, uses a ZKP-friendly commitment scheme with a linear structure, and is compatible with the IT-MACs in authenticated secret sharing. Meanwhile, our approach achieves high statistical security, while \cite{peng2025reliable}'s check fails with a non-negligible probability.

\subsubsection{\textbf{Micro-benchmarks of MPC building blocks.}}

Table~\ref{tab:microbench} compares the performance of our fundamental MPC backend across core components against \textsc{Shark}~\cite{gupta2025shark}. Each experiment is executed one million times. Our framework achieves substantial improvements in both computation and communication efficiency. Specifically, our optimized DPF and DCF designs reduce communication by {1.57}$\times$ and execution time by {4.91}$\times$. For DReLU, our protocol requires only {19.3\%} of the running time and {70.5\%} of the communication compared with \textsc{Shark}. For Truncation, the corresponding figures are {20.6\%} of the running time and {67.2\%} of the communication. For LUT, our implementation achieves {58.1\%} of the running time and {91.0\%} of the communication cost relative to \textsc{Shark}. The communication computational efficiency improvements primarily result from the reduced size of DPF and DCF keys, while the computational efficiency gains arise from fewer AES calls, which dominate the runtime cost of FSS-based evaluation.

\begin{table}[htbp]
\centering
\caption{Comparison on a microbench for MPC building blocks between \method{} and Shark~\cite{gupta2025shark}.}
\vspace{-5pt}
\resizebox{0.95\columnwidth}{!}{%
\setlength{\tabcolsep}{13pt}
\begin{tabular}{cccc}
\toprule
Function               & Protocol                      & Comm. (MB)                  & Run time (s)            \\ \hline
\multirow{2}{*}{DPF/DCF} & \method{}                 & 2080                    & 0.45                     \\ 
                       & Shark~\cite{gupta2025shark}                   & 3266                     &  2.21                      \\ \hline
\multirow{2}{*}{DReLU} & \method{}                 & 2178                    & 1.79                     \\ 
                       & Shark~\cite{gupta2025shark}                   & 3090                     &  9.28                      \\ \hline
\multirow{2}{*}{Truncation} & \method{}                 & 2856                     & 2.39                     \\ 
                       & Shark~\cite{gupta2025shark}                & 4249                     & 11.58                     \\ \hline
\multirow{2}{*}{LUT} & \method{}                 & 416                     & 4.79                     \\ 
                       & Shark~\cite{gupta2025shark}                & 457                     & 8.25                     \\
\bottomrule
\end{tabular}
}
\label{tab:microbench}
\end{table}

\begin{figure}[h]
    \centering
    \includegraphics[width=0.45\textwidth]{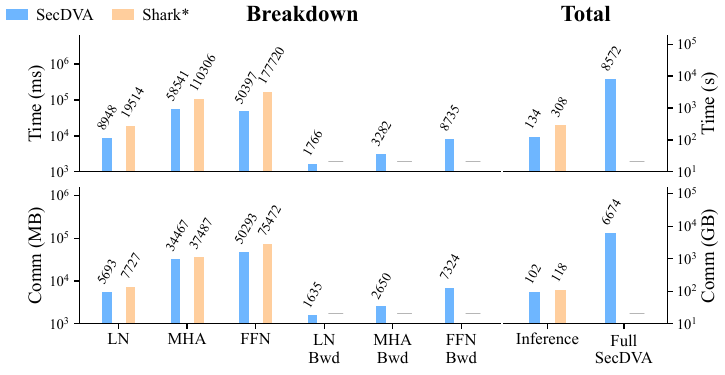}
    \vspace{-5pt}
    \caption{End-to-end performance of \method{}. Shark* denotes our re-implementation of Shark~\cite{gupta2025shark} with higher precision, ``Bwd'' denotes the backward differentiation, and ``---'' indicates that the result is not available.}
    \label{fig:end2end}
    \vspace{-10pt}
\end{figure}

\subsubsection{\textbf{Benchmark of Transformer-based MPC protocols.}}
Table~\ref{tab:protocol} reports the evaluation of Transformer-based MPC protocols, including LayerNorm, Softmax, and GeLU. The matrix inverse is additionally considered for NTK-relevant operations in \method{}. Since Shark does not implement LayerNorm or matrix inverse, we denote the results as ``--''. We also provide the mean error of each protocol under a representative input range (i.e., the CIFAR-100 dataset, whose inputs fall between $-10$ and $10$, and without losing generality, other datasets fall in this range after standardization that is commonly used in ML training) to results calculated in plain text with float precision.

For Softmax, our protocol requires $1.95\times$ the communication and $1.66\times$ the running time of Shark. However, our construction achieves a mean error of only $0.0001$, compared to $0.0051$ for Shark. For GeLU, our protocol incurs $5.30\times$ the communication and $2.36\times$ the running time of Shark. This is because Shark employs a coarse approximation using a quadratic polynomial, while we adopt a piecewise 6th-degree polynomial. On random inputs within $[-10, 10]$, Shark produces a mean error of $2.3749$, which deviates significantly from the ground truth. Since our protocols are designed for precision-sensitive applications, and intermediate results must be preserved to compute derivatives, it is essential to prioritize the precision of results and sacrifice efficiency.

For a fair comparison, we also adapt Shark into Shark* with the same precision as \method{}. Under this setting, \method{} shows clear efficiency advantages: for LayerNorm, it requires only {68.7\%} of the communication and {28.5\%} of the running time of Shark*; for Softmax, only {86.3\%} of the communication and {19.1\%} of the running time; and for GeLU, only {66.0\%} of the communication and {26.4\%} of the running time. These results present the benefits of \method{} in precision-sensitive settings, and most of the improvements stem from our efficient constructions of DReLU and truncation.

For matrix inversion, we found that 30 iterations of the Newton method are required to obtain a reasonable approximation of the matrix inverse, during which the error grows rapidly. We temporarily double the precision (e.g., to 24 bits) when computing the inverse, and then shift it back once the iterations are completed. Finally, our inverse for a matrix of 32$\times$32 only costs 212 MB and 1.27 seconds with a mean error of 0.0009.

\begin{table}[htbp]
\centering
\caption{Comparison on a benchmark of transformer-based MPC protocols between \method{} and Shark~\cite{gupta2025shark}. Shark* denotes our re-implementation of Shark with higher precision.}
\label{tab:protocol}
\vspace{-5pt}
\resizebox{\columnwidth}{!}{%
\setlength{\tabcolsep}{3pt}
\begin{tabular}{cccccc}
\hline
Function                        & Size                & Protocol                    & Comm. (MB) & Run time (s) & Mean error \\ \hline
\multirow{3}{*}{LayerNorm}      & \multirow{3}{*}{197$\times$192} & \method                     & 228        & 0.45               & 0.0081     \\
                                &                     & Shark~\cite{gupta2025shark} & -          & -                  & -          \\
                                &                     & Shark*                      & 332          & 1.58                   & 0.0081           \\ \hline
\multirow{3}{*}{Softmax}        & \multirow{3}{*}{197$\times$197} & \method                     & 564        & 1.15               & 0.0001     \\
                                &                     & Shark~\cite{gupta2025shark} & 288        & 0.69               & 0.0051     \\
                                &                     & Shark*                      & 653           &  6.03                  & 0.0001           \\ \hline
\multirow{3}{*}{GeLU}           & \multirow{3}{*}{197$\times$768}   & \method                     & 3620       & 3.36               & 0.0013     \\
                                &                     & Shark~\cite{gupta2025shark} & 682        & 1.42               & 2.3749     \\
                                &                     & Shark*                      & 5488           & 12.72                   & 0.0013           \\ \hline
\multirow{3}{*}{Matrix Inverse} & \multirow{3}{*}{32$\times$32}   & \method                     & 212        & 1.27               & 0.0009     \\
                                &                     & Shark~\cite{gupta2025shark} & -          & -                  & -          \\
                                &                     & Shark*                      & -           &  -                  & -           \\ \hline
\end{tabular}%
}
\vspace{-10pt}
\end{table}

\subsubsection{\textbf{End-to-end Benchmark.}} 
Figure~\ref{fig:end2end} presents the end-to-end benchmark of \method{} for evaluating a ViT-Tiny model. 
Since the original Shark implementation exhibits poor precision, we instead compare Shark* with \method{}. Moreover, as Shark does not support differentiation, we exclude this part from the comparison. Our protocols demonstrate significant improvements in running time: for the entire inference, \method{} requires only {43.6\%} of the running time of \textsc{Shark*}. The majority of this gain is attributed to the protocols listed in Table~\ref{tab:protocol}. The advantage of end-to-end is not as significant as the improvement brought by LayerNorm, Softmax, and GeLU, since other operations, such as matrix multiplication, are also involved in the inference. \method{} also outperforms \textsc{Shark*} in communication efficiency, requiring only {86.9\%} of the communication overhead. For the implementation of the mini-batch model parameter displacement calculation with the landmark correction, we conduct experiments on the CIFAR-100 dataset (for other datasets, the calculation is the same if we apply the same standardization). \method{} requires approximately 8,572 seconds and 6,674~GB of communication to compute the parameter displacement, validating the practical feasibility in the real world, especially with large potential in incorporating implementation optimizations such as computation parallelization and GPU acceleration. The outputs also retain relatively high precision, with a cosine similarity of {0.933} and Euclidean distance of {0.04} to plaintext results.


\section{Conclusion}
We present \method{}, the first maliciously secure MPC framework to assess dataset value in modern machine learning data markets. To make \method{} efficient, we use NTK theory to approximate model parameter changes. \method{} employs function secret sharing for active security and optimizes key MPC primitives, including distributed point and comparison functions, and transformer components. It also pioneers efficient and precise MPC implementations for matrix inversion and end-to-end gradient computation.

\section*{Acknowledgments}
This research is supported by the Singapore Ministry of Education Tier 1 Grant (RG19/25) and the NTU–NAP Startup Grant (024584-00001). We also thank the anonymous reviewers for their valuable suggestions. We used AI only for editorial purposes in this paper, such as grammar checking, and we inspected all outputs to ensure originality.

\bibliographystyle{IEEEtran}
\bibliography{sample}

\end{document}